\documentclass[12pt]{article} 
\usepackage{amsmath,amssymb}  %

\usepackage{pst-all,citesort}    
\usepackage[square,authoryear]{natbib}

\def\cocoa{{\hbox{\rm C\kern-.13em o\kern-.07em C\kern-.13em o\kern-.15em A}}}
\newcounter{definicion}
\newcounter{ejemplo}

\newtheorem{definition}[definicion]{Definition}
\newtheorem{example}[ejemplo]{Example}

\title{Smooth supersaturated models}

\author{Ron A. Bates\footnote{Department of Statistics, London School of Economics, London WC2A 2AE, UK},
Hugo Maruri-Aguilar$^*$\footnote{Email address: \texttt{H.Maruri-Aguilar@lse.ac.uk}}
, Henry P. Wynn$^*$}

\begin{document}
\maketitle

\begin{abstract}
In areas  such as kernel smoothing and non-parametric regression
there is emphasis on smooth interpolation and smooth statistical
models. Splines are known to have optimal smoothness properties in
one and higher dimensions. It is shown, with special attention to
polynomial models, that smooth interpolators can be constructed by
first extending the monomial basis and then minimising a measure of
smoothness with respect to the free parameters in the extended
basis. Algebraic methods are a help in choosing the extended basis
which can also be found as a saturated basis for an extended
experimental design with dummy design points. One can get
arbitrarily close to optimal smoothing for any dimension and over
any region, giving a simple alternative models of spline type. The
relationship to splines is shown in one and two dimensions. A case
study is given which includes benchmarking against kriging methods.
\end{abstract}



\section{Introduction}\label{sec_intro}
There is a considerable literature on smooth interpolation and its
statistical counterparts. The area of non-parametric regression is
an example. The optimal smoothness properties of splines have a
substantial literature. The optimality result for one dimensions is
attributed to \cite{H1957} and for two dimensions, where thin-plate
splines are optimal, to \cite{D1976}; see \cite{M2002} for a nice
review on spline optimality and \cite{KW1970} for an overview. In
computer experiments Bayesian kriging using Gaussian kernel
stochastic process models has been preferred to splines,
\cite{SWMW1989}, \cite{KH2001}, and have also become popular in
machine learning: \cite{RW2005}. Of course, the connection between
kriging and spline is thoroughly researched and, for example,
splines can arise as kriging (conditional expectation) interpolators
for special Gaussian stochastic processes: \cite{KW1970}.

Raw polynomial interpolation is known in general not to have optimal
rates of interpolation unless special sampling (design) points are
used such as in Tchebychev approximation. On the other hand the {\em
existence} of polynomial interpolators over an arbitrary design is
at the core of the newer theory of algebraic statistics: for any
arbitrary design in $d$ dimensions there is always a monomial basis
out of which we can build a polynomial interpolator. This was
introduced into statistics by \cite{PW1996}, covered at length in
the monograph \cite{PRW2000} and was also the basis for
\cite{BGW2003} which can be seen as the forerunner of the present
paper.

The aim of the present paper is to try to have the best of both
worlds: to draw a little on the algebraic theory but principally to
show, in an rather elementary way, how to construct smooth
polynomial interpolators or statistical models. This is achieved by
extending the model basis and using this freedom to optimise a
measure of smoothness.  It should be pointed out that the use of
polynomials to build kernels with pre-specified properties is
familiar in signal processing, see \cite{LXW2004}. By extending the
model basis we can show that our interpolators get arbitrarily close
to optimal interpolators, which are typically in the spline family.

\subsection{Monomial bases and extended bases}
Recent work in the area of ``algebraic statistics" shows how to
construct estimable (identifiable) monomial bases for polynomial
regression and we start with a very short description. Having said
this, it is not necessary to use these methods, nor indeed to use
polynomials. For example a Fourier (trigonometric) basis may be
used. The point is that we shall need an extended basis with certain
conditions and the algebra is one way of achieving this.

We start with a set of factors $x=(x_1,\ldots, x_d)$. For a set of
nonnegative integers $\alpha=(\alpha_1,\ldots,\alpha_d)$, a
monomial, such as $x_1^2x_2$, is written
$x^\alpha=x_1^{\alpha_1}\cdots x_d^{\alpha_d}$, and a polynomial is
a linear combination of
monomials. 
A design $D_n$ is a set of $n$ distinct points in $d$ dimensions,
$D_n=\{x^{(1)},\ldots,x^{(n)}\}$, $x^{(i)}\in\mathbb
R^d,i=1,\ldots,n$.

The algebraic methods give us the following: {\em given an
experimental design}, $D_n$, {\em it is always possible to find a
saturated non-singular monomial basis} $B_L=\{x_{\alpha}, \alpha \in
L \}$. Thus, the size of the basis is equal to the size of the
design $|L|=|D_n| =n$ and the $n \times n$ $X$-matrix, $X = \{
x^{\alpha}\}_{x \in D_n, \alpha \in L}$ is non-singular. We call
such a basis a {\em good saturated basis} for the design. The
intuition behind algebraic methods is simple: terms are included in
the good saturated basis according to a term ordering and a rank
inclusion criterion. For details on term orderings see \cite{CLO96},
and for description of the algebraic technology see \cite{PRW2000}.

\begin{example} \label{ex_sobol}{\normalfont Let $D_{24}$
to be the first $24$ points of a bidimensional Sobol's space filling
sequence. An implementation of the description of Sobol' sequence by
\cite{BF1988} is available in the language R, see \cite{IG1996}.
Then by selecting terms with a degree lexicographic term order
$x_1\succ x_2$, a good saturated basis with $24$ monomials is
identified for $D_{24}$. This model includes the monomials
$x_2^6,x_1x_2^5,x_1^2x_2^4$ plus all the terms of a model of total
degree five. This basis will be extended in the example of Section
\ref{ex2}.}\end{example}

It will be critical in our development that we may extend a basis.
By this we mean we keep the design $D_n$ fixed but take a larger set
of $N
>n$ monomials, hence the term ``supersaturated" in the title if the
paper. But we require a condition contained in the following
definition.


\begin{definition}Given a design $D_n$, with sample size $n$,
a {\em good  supersaturated basis} is a basis $B_M=\{x^{\alpha},
\alpha \in M \}$ with $|B| = N >n$ such that there is a hierarchical
non-singular sub-basis of size $n$. \end{definition}

Here is an example to show that we have to be a little careful. Let
us start with a rather poor design in two dimensions: $D_4=\{ (0,0),
(1,1), (2,2),(3,3)\}$. Then, and this is obvious without any
algebra, there are only two good saturated model bases $\{1, x_1,
x_1^2, x_1^3\}$ or $\{1, x_2, x_2^2, x_2^3\}$. From this we can see
that the extended basis $\{1, x_1, x_1^2, x_2, x_2^2\}$ with five
terms is not good as there is no good sub-basis of size four.

If we start with a non-singular basis for a design $D_n$ and extend
it, in any way, then we always obtain a good supersaturated basis.
But there is a revealing way of generating a good supersaturated
basis and that is by extending the design $D_n$ to a design $D_N$
with $N$ points and finding a good saturated basis for larger
design, which contains the good basis for $D_n$. The algebra shows
that this is always possible. This leads to a second, and
equivalent, way of producing the smooth models which will be called
the ``dummy design" method, covered in sub-section 2.2.

\section{Smooth interpolators}\label{sec_smooth}
The basic idea of this paper may seem at first to be somewhat
contradictory. We start with given polynomial interpolator and by
extending the basis make the interpolator smoother. Although one may
naturally associates higher order polynomial terms with lack of
smoothness, we can, in fact, extend the basis and use the freedom
this gives to {\em increase} smoothness.

Let the experimental design be $D_n$ and $y_1, \ldots, y_n$ be real
values (observations) at the design points $x^{(i)}\in
D_n,i=1,\ldots,n$, respectively. Let $B_M$ be a good supersaturated
basis for the design $D_n$ and let
\begin{equation}
y(x) = \sum_{\alpha \in M} \theta_{\alpha} x^{\alpha}
\end{equation}
be a polynomial in that basis. A good supersaturated model will be
sought for using a measure of smoothness.


In one dimension ($d=1$) we shall adopt the following measure of
smoothness based on the second derivative
\begin{equation}\label{hw_curvature}\Psi_2=\int_{\mathcal X}| y''(x)|^2dx,\end{equation}
where the integration is carried out in a desired region $\mathcal
X\subset\mathbb R$. For higher dimensions the Hessian is
$$H(y(x))=\left\{\frac{\partial^2 y(x)}{\partial x_i\partial x_j }\right\},$$
and we have \begin{equation}\label{hw_multicurvature}\sum_{ij}
\left(\frac{\partial^2 y(x)}{\partial x_i\partial
x_j}\right)^2=||H(y(x))||^2=\mbox{trace}\left(H(y(x))^2 \right).
\end{equation}
Then define
\begin{equation}
\Psi_2 = \int_{\mathcal X}||H(y(x))||^2 dx,
\end{equation}
for some desired region $\mathcal X \subset \mathbb R^d$.

The smooth interpolator is $\hat y(x) = \sum_{\alpha \in M}
\hat\theta_{\alpha} x^{\alpha}$, where the coefficients $
\hat\theta_\alpha$ are selected to minimise smoothness subject to
the interpolation condition, i.e. solving the constrained
optimisation problem
\begin{equation}
\min_{\theta} \Psi_2(y(x))\;\;\mbox{subject to}\;\;y_i =
\hat{y}(x^{(i)}),\; i=1,\ldots,n
\end{equation}
In the next subsection we give the solution of this constrained
problem and the in the second subsection give the dummy design
method, which is equivalent.
\subsection{The constrained problem}
The only technical difficulty arises from the fact that linear parts
of the model make no difference to the criterion $\Psi_2$ but do
affect the interpolation. It is necessary to partition the
$X$-matrix to take account of this.

Let $f(x)$ and $\theta$ respectively be the vectors which hold the
good supersaturated basis and the parameters so that we can write
(1) as $y(x) = \theta^T f(x)$. Denote
\mbox{$f^{(ij)}=\frac{\partial^2f(x)}{\partial x_i\partial
x_j}$ }
 and define
\begin{equation}\label{ec_K}K=\int_\mathcal
X\left(\sum_{i,j=1}^{k}f^{(ij)}{f^{(ij)}}^T\right)dx,
\end{equation}
Then we see that
\begin{equation}
\Psi_2(y(x))  =  \theta^T K \theta
\end{equation}
The technical difficulty mention above means that $K$ may not be
full rank. In particular any linear term in the models basis will
give zero entries. Call this entries {\em structural zeros}. Permute
the rows and columns of $K$ so that the structural zeros are
adjacent:
\begin{equation}\label{ec_partk}
K = \left[
\begin{array}{r  r}
0 & \;\;\;\;\;0 \\
0 & \;\;\;\;\;\tilde{K}
\end{array}
\right]
\end{equation}
\vspace{3mm} Let $X = [X_0, X_1]$, $f=(f_0^T: f_1^T)^T$ and $\theta
= (\theta_0^T: \theta_1^T)^T$ be the corresponding rearranged and
partitioned versions of $X_n$, $f$ and $\theta$, respectively. The
matrix $X$ has $n$ rows and as many columns as terms in $f$. Let
$y$, be the column vector with $n$ observations and note that
$\Psi_2 = \theta_1^T \tilde{K}\theta_1$.

With this the constrained quadratic problem (5) is:
\begin{equation}
\min_{\theta}\; \theta_1^T \tilde{K} \theta_1 \;\;\;\mbox{subject
to}\;\;\;X_0 \theta_0 + X_1 \theta_1 = y
\end{equation}
Let $2\lambda$ be an $n \times 1$ vector of Lagrange multipliers
($2$ is for convenience) so that the Lagrangian is
$$\theta_1^T \tilde{K} \theta_1 - 2\lambda(X_0 \theta_0 + X_1
\theta_1). $$ After differentiation the full set of equations for
$\theta_0, \theta_1$ and $\lambda$ can be written in block form
\begin{equation}\label{ec_totalA} \left[
\begin{array}{llr}
X_0\;\; & X_1  & 0 \\
0\;\;  &  \tilde{K} & -X_1^T \\
0\;\; & 0 & X_0^T
\end{array}
\right] \left[
\begin{array}{lll}
\theta_0 \\
\theta_1\\
\lambda
\end{array}
\right] = \left[
\begin{array}{lll}
y \\
0\\
0
\end{array}
\right]
\end{equation}
If the matrix on the left hand side is nonsingular we obtain a
unique solution $\hat{\theta}_0, \hat{\theta}_1, \hat{\lambda}$. The
following three conditions guarantee this.

(i) The full basis is a good supersaturated basis for $D_n$ so that
X is full rank.

(ii) $X_0$ is full rank.

(iii) $\tilde{K}$ is full rank and thus invertible.

The full matrix inverse with solutions $\hat{\theta}_0,
\hat{\theta_1}, \hat{\lambda}$ are given in Appendix 1. Finally,
using these results, we express the smooth estimator as
$$\hat{y}(x) = \hat{\theta}_0 f_0 + \hat{\theta}_1 f_1 =
\hat{\theta}f(x)$$ and the optimal $\Psi_2$ as
$$\Psi_2^*= \hat{\theta}_1^T \tilde{K}\hat{\theta_1}.$$
In applications, as is common with quadratic programme, we simply
invert the matrix on the right hand side of (9) using a fast
numerical method. Thus, given the design $D_n$, the good
supersaturated basis and $\tilde{K}$, the method is fairly
straightforward to implement.

It is revealing to consider the case where $K$ is nonsingular. Then
we do not need the partition of Equation (\ref{ec_partk}) and
instead can write Equation (\ref{ec_totalA}) as
$$
\left[
\begin{array}{lr}
X  & 0 \\
\tilde{K} & -X
\end{array}
\right] \left[
\begin{array}{lll}
\theta\\
\lambda
\end{array}
\right] = \left[
\begin{array}{lll}
y \\
0
\end{array}
\right]
$$
Which has the solution:
$$
\hat{\theta} = (X^TX + K(I-P)K)^{-1} X^Ty
$$
where $P= X^T(XX^T)^{-1}X$ is the projector onto the row space of
$X$. Thus, although $X^TX$ is not invertible, because we have a
supersaturated model, the second term $K(I-P)K$ on the left hand
side can be seen as a smoothness induced regularisation of the
problem which compensates for this singularity.

\subsection{The dummy design method}\label{sec_smooth}
For simplicity of development we assume that $K$ is non-singular in
the present case. Let $D_N$ be a large design, with $N>n$ distinct
points, which contains the original design $D_n$ and write
$$D_N= D_n \cup D_q,$$
where $q=N-n$. Let $h(x)$ be a good saturated basis for $D_n$, and
let $f(x)$ be an (extended) good saturated basis for $D_N$,
$f(x) = (h(x)^T, g(x)^T)^T$. 
Also extend the observation vector to $z = (y^T,z^T)^T$ where, as
before $y$ holds the ``true" observations taken at points in $D_n$,
and $z$ can be thought of as dummy observations on the design $D_q$.
The extended model we write
\begin{equation}
y(x)=f(x)^T\theta=h^T(x)\beta+g^T(x)\gamma
\end{equation}
and assume, as in the last section, that $y(x)$ interpolates the
observations $y$ over $D_n$.

We now minimize $\Psi_2$ over the the choice of dummy observations
$z$ which is now an unconstrained optimization problem, but with a
reduced set of free parameters, namely $z$. The constrained
optimization (8) and this unconstrained optimization are equivalent
in the case that the full basis is  a good for the full design,
$D_N$. This is because of the one to one correspondence between
observations and parameters and the fact that the interpolation
constraint is the same in both cases.

The unconstrained problem is:
\begin{equation}
\min_z \; (y^T:z^T) {X_N^{-1}}^T K X_N^{-1}\; \mbox{$ y
\choose{z}$}.
\end{equation}

Where $X_N$ is the $X$-matrix for the full large model $f(x)$.
First, let the following matrix be partitioned according to the
model bases $f(x) = (h(x)^T, g(x)^T)^T$:
\[A={X_N^{-1}}^TKX_N^{-1}=\left(\begin{array}{cc}A_{11}&A_{12}\\A_{21}&A_{22}\end{array}\right).\]
Then after expanding (11) and differentiating, the optimal $z$ is
$$\hat{z} = -A_{22}^{-1}A_{21}y$$
and the minimum value of the smoothness is
$$\Psi_2^* = y^T Q \; y,$$
where $Q = A_{11}- A_{22}^{-1}A_{21}$. The smooth interpolator is
\begin{equation}
\hat{y}(x) = f^T(x)X_N^{-1}\mbox{$ y \choose{\hat{z}} $} =
f^T(x)X_N^{-1}\mbox{$ I \choose{-A_{22}^{-1}A_{21}} $}y=
f^T(x)K^{-1} (X_{11} : X_{12}) Q y
\end{equation}
where
\[X_N=\left(\begin{array}{cc}X_{11}&X_{12}\\X_{21}&X_{22}\end{array}\right)\]
is the appropriate partition of $X_N$, i.e. the rows of $X_N$ are
indexed by $D_n$ and $D_q$, while the columns are indexed by $h(x)$
and $g(x)$.

The last equality and the equivalence to the solution in the last
subsection is shown for the case that $K$ is non-singular. The
equivalence in general holds under conditions (i), (ii) and (iii) in
that section. We not that the solution does do not depend on the
dummy design $D_q$, except in so far as it is involved in
guaranteeing that we have a good supersaturated basis.

\section{One and two dimensions}
\subsection{A one dimensional example: spline-like behavior}
\label{ex1}

In this example, smooth saturated models are used for interpolating
a known univariate function. The function considered is the sine
cardinal
\[m(x)=\mbox{sinc}(ax+b)=\sin(ax+b)/(ax+b)\] with $a=15\pi/2$ and
$b=-10\pi/2$. The region over which the interpolators will be
smoothed is $\mathcal X=[0,1]$.

\begin{figure}[h!]  
\begin{center}
\psset{unit=38mm,linewidth=0.85pt} 
 \begin{pspicture}(0,-0.5)(1.1,1.25) 
     \put(0,0){}

\psset{linecolor=black}

\psline[linestyle=dashed] (0,0) (0.0125,0.0922773127852348)
(0.025,0.159713639540717) (0.0375,0.205457979354918)
(0.05,0.232453254751687) (0.0625,0.243442070732292)
(0.075,0.240972473817463) (0.0875,0.227403711089436)
(0.1,0.204911989234) (0.1125,0.175496233582533)
(0.125,0.140983847154052) (0.1375,0.103036469697255)
(0.15,0.0631557367325623) (0.1625,0.022689038594161)
(0.175,-0.0171647205279515) (0.1875,-0.0553493635459215)
(0.2,-0.0909456814320001) (0.2125,-0.123165674176497)
(0.225,-0.151346791745736) (0.2375,-0.174946175040017)
(0.25,-0.193534896851563) (0.2625,-0.206792202822486)
(0.275,-0.214499752402739) (0.2875,-0.216535859808075)
(0.3,-0.212869734978) (0.3125,-0.20355572453373)
(0.325,-0.188727552736152) (0.3375,-0.168592562443772)
(0.35,-0.143425956070686) (0.3625,-0.113565036544524)
(0.375,-0.0794034482644048) (0.3875,-0.0413854180588999) (0.4,0)
(0.4125,0.0442247029189615) (0.425,0.0907272592653121)
(0.4375,0.138918706769113) (0.45,0.18818829208519)
(0.4625,0.237909233691178) (0.475,0.287444480929556)
(0.4875,0.336152473049722) (0.5,0.38339289825)
(0.5125,0.428532452719693) (0.525,0.470950599681144)
(0.5375,0.510045328431762) (0.55,0.545238913386072)
(0.5625,0.575983673117733) (0.575,0.601767729401642)
(0.5875,0.622120766255934) (0.6,0.636619788984003)
(0.6125,0.644894883216608) (0.625,0.646634973953869)
(0.6375,0.641593584607286) (0.65,0.629594596041937)
(0.6625,0.610538005618277) (0.675,0.584405686234362)
(0.6875,0.551267145367849) (0.7,0.511285284118014)
(0.712499999999999,0.464722156247824)
(0.724999999999999,0.411944727225947)
(0.737499999999999,0.35343063326885)
(0.749999999999999,0.289773940382815)
(0.762499999999999,0.221690903405928)
(0.774999999999999,0.150025725050192)
(0.787499999999999,0.0757563149435754) (0.799999999999999,0)
(0.812499999999999,-0.0759804731785749)
(0.824999999999999,-0.150771665980095)
(0.837499999999999,-0.222803202020444)
(0.849999999999999,-0.290342251461269)
(0.862499999999999,-0.351487723296216)
(0.874999999999999,-0.404164506308277)
(0.887499999999999,-0.446117710028744)
(0.899999999999999,-0.474906905693985)
(0.912499999999999,-0.48790036720446)
(0.924999999999999,-0.482269312081755)
(0.937499999999999,-0.454982142427511)
(0.949999999999999,-0.402798685880427)
(0.962499999999999,-0.322264436574955)
(0.974999999999999,-0.209704796098777)
(0.987499999999998,-0.0612193144509661) (1,0.127324069000025)

\psline[linecolor=black] (0,0) (0.0125,0.122128026685379)
(0.025,0.210995016817803) (0.0375,0.271062225568747)
(0.05,0.306456890899128) (0.0625,0.320985402291265)
(0.075,0.318146271889113) (0.0875,0.301142908046754)
(0.1,0.27289619128517) (0.1125,0.236056852657274)
(0.125,0.193017654521217) (0.1375,0.145925373721962)
(0.15,0.0966925871811238) (0.1625,0.0470092598950829)
(0.175,-0.00164586465863648) (0.1875,-0.0479940717067144)
(0.2,-0.0909456799571207) (0.2125,-0.129589243967926)
(0.225,-0.163180954107843) (0.2375,-0.191134234108513)
(0.25,-0.213009536208497) (0.2625,-0.228504333889021)
(0.275,-0.237443312201454) (0.2875,-0.239768755686484)
(0.3,-0.235531133885073) (0.3125,-0.224879884441101)
(0.325,-0.208054393795775) (0.3375,-0.185375175473735)
(0.35,-0.157235245960933) (0.3625,-0.124091698174194)
(0.375,-0.0864574725225446) (0.3875,-0.0448933255602731) (0.4,0)
(0.4125,0.0475894302923703) (0.425,0.0972169611862687)
(0.4375,0.148206936949588) (0.45,0.199873074795512)
(0.4625,0.251525314447281) (0.475,0.302476466343099)
(0.4875,0.352048662249199) (0.5,0.399579608281241)
(0.5125,0.444428640333916) (0.525,0.485982581918919)
(0.5375,0.523661404411015) (0.55,0.556923689702648)
(0.5625,0.585271895266483) (0.575,0.608257421626535)
(0.5875,0.625485482237357) (0.6,0.636619775771505)
(0.6125,0.64138696081547) (0.625,0.639580932973677)
(0.6375,0.63106690438064) (0.65,0.615785285621952)
(0.6625,0.59375537006273) (0.675,0.565078820585019)
(0.6875,0.529942958733169) (0.7,0.488623856267324)
(0.712499999999999,0.441489229125589)
(0.724999999999999,0.389001133794201)
(0.737499999999999,0.331718466085842)
(0.749999999999999,0.270299262326635)
(0.762499999999999,0.205502802950889)
(0.774999999999999,0.138191518504652)
(0.787499999999999,0.0693326980571314) (0.799999999999999,0)
(0.812499999999999,-0.0686252346220719)
(0.824999999999999,-0.135252866679732)
(0.837499999999999,-0.198483040709615)
(0.849999999999999,-0.256805464564039)
(0.862499999999999,-0.308598886528909)
(0.874999999999999,-0.352130770054522)
(0.887499999999999,-0.385557166078407)
(0.899999999999999,-0.406922782938054)
(0.912499999999999,-0.414161253877978)
(0.924999999999999,-0.405095602146197)
(0.937499999999999,-0.377438903684308)
(0.949999999999999,-0.328795147407803)
(0.962499999999999,-0.256660293079108)
(0.974999999999999,-0.158423526771813)
(0.987499999999998,-0.0313687139267671) (1,0.127323949999905)

\psline[linecolor=black] (0,0) (0.0125,0.0105515094908516)
(0.025,0.0242294763207508) (0.0375,0.0384207490682549)
(0.05,0.0511226660281489) (0.0625,0.0608700583992138)
(0.075,0.0666669606881647) (0.0875,0.0679228957673819)
(0.1,0.06439360202406) (0.1125,0.0561260700383989)
(0.125,0.0434077562284615) (0.1375,0.0267198408993214)
(0.15,0.00669439813412654) (0.1625,-0.0159246550352978)
(0.175,-0.0403169627406841) (0.1875,-0.0656176202347675)
(0.2,-0.09094568225632) (0.2125,-0.115429422365147)
(0.225,-0.138228475809421) (0.2375,-0.158552998487741)
(0.25,-0.175679974568262) (0.2625,-0.188966805327311)
(0.275,-0.197862311769835) (0.2875,-0.201915283594065)
(0.3,-0.200780707062778) (0.3125,-0.194223804343532)
(0.325,-0.182122016880224) (0.3375,-0.1644650653584)
(0.35,-0.141353218826644) (0.3625,-0.112993905536422)
(0.375,-0.0796967980628129) (0.3875,-0.0418675052684274) (0.4,0)
(0.4125,0.0453320592093573) (0.425,0.0934850570018413)
(0.4375,0.143755939504801) (0.45,0.195393518079861)
(0.4625,0.247610348816788) (0.475,0.299595080920497)
(0.4875,0.350525137755778) (0.5,0.399579597987515)
(0.5125,0.445952144253752) (0.525,0.488863946809627)
(0.5375,0.527576349579386) (0.55,0.561403226054411)
(0.5625,0.58972287247467) (0.575,0.611989305731367)
(0.5875,0.627742833428446) (0.6,0.63661976354018)
(0.6125,0.638361121103209) (0.625,0.632820239379761)
(0.6375,0.619969092930745) (0.65,0.599903240034667)
(0.6625,0.572845241892722) (0.675,0.539146426053911)
(0.6875,0.499286861502007) (0.7,0.453873412838654)
(0.712499999999999,0.403635741001253)
(0.724999999999999,0.349420117954786)
(0.737499999999999,0.292180922792554)
(0.749999999999999,0.232969686686005)
(0.762499999999999,0.172921554118048)
(0.774999999999999,0.113239027842667)
(0.787499999999999,0.0551728650022909) (0.799999999999999,0)
(0.812499999999999,-0.0510016955221033)
(0.824999999999999,-0.0965817769076693)
(0.837499999999999,-0.135549132896273)
(0.849999999999999,-0.166807281363077)
(0.862499999999999,-0.189393358200135)
(0.874999999999999,-0.202520874820095)
(0.887499999999999,-0.205626384994105)
(0.899999999999999,-0.198420193597883)
(0.912499999999999,-0.180941239811887)
(0.924999999999999,-0.153616287355135)
(0.937499999999999,-0.117323554298764)
(0.949999999999999,-0.0734609150380265)
(0.962499999999999,-0.0240188069691385)
(0.974999999999999,0.0283420255537976)
(0.987499999999998,0.0802078174256167) (1,0.127323966600329)

\psline[linecolor=black] (0,0) (0.0125,0.0124773232845559)
(0.025,0.0290147091062953) (0.0375,0.0463827223468966)
(0.05,0.062124497944711) (0.0625,0.0744572396532906)
(0.075,0.0821809691145355) (0.0875,0.0845942416316231)
(0.1,0.081416549165309) (0.1125,0.072717135215611)
(0.125,0.0588499503893118) (0.1375,0.0403944815921372)
(0.15,0.0181021919228962) (0.1625,-0.00715168751471531)
(0.175,-0.0344112685285136) (0.1875,-0.062678269989882)
(0.2,-0.0909456815904959) (0.2125,-0.118227069434326)
(0.225,-0.143581761557092) (0.2375,-0.166136147326922)
(0.25,-0.18510132054154) (0.2625,-0.199787291898901)
(0.275,-0.209613992379737) (0.2875,-0.214119284942088)
(0.3,-0.212964197789449) (0.3125,-0.205935588335739)
(0.325,-0.192946442851882) (0.3375,-0.174034012640369)
(0.35,-0.149355983445752) (0.3625,-0.119184870670523)
(0.375,-0.0839008288276113) (0.3875,-0.0439830595219864) (0.4,0)
(0.4125,0.0474015449074059) (0.425,0.0975081160013806)
(0.4375,0.149552056300793) (0.45,0.202724481448182)
(0.4625,0.256188644042834) (0.475,0.309093523387503)
(0.4875,0.360587491347917) (0.5,0.40983190727812)
(0.5125,0.456014499103006) (0.525,0.498362391788509)
(0.5375,0.536154648567431) (0.55,0.568734194427987)
(0.5625,0.595518995510089) (0.575,0.616012372193429)
(0.5875,0.629812327799043) (0.6,0.636619778965338)
(0.6125,0.636245577896786) (0.625,0.628616220823725)
(0.6375,0.613778141148489) (0.65,0.591900489891308)
(0.6625,0.563276310190648) (0.675,0.528322016746655)
(0.6875,0.487575095239318) (0.7,0.441689940888006)
(0.712499999999999,0.391431759458206)
(0.724999999999999,0.337668458162955)
(0.737499999999999,0.281360458038678)
(0.749999999999999,0.223548363518912)
(0.762499999999999,0.165338429065752)
(0.774999999999999,0.107885766858857)
(0.787499999999999,0.0523752436742519) (0.799999999999999,0)
(0.812499999999999,-0.0480623175592854)
(0.824999999999999,-0.0906760539655593)
(0.837499999999999,-0.126776135608893)
(0.849999999999999,-0.155399456737634)
(0.862499999999999,-0.175718685559367)
(0.874999999999999,-0.187078647548638)
(0.887499999999999,-0.189035285482277)
(0.899999999999999,-0.181397210824869)
(0.912499999999999,-0.164269856933842)
(0.924999999999999,-0.138102240434179)
(0.937499999999999,-0.103736332957752)
(0.949999999999999,-0.0624590413146251)
(0.962499999999999,-0.0160567900196185)
(0.974999999999999,0.0331273040340534)
(0.987499999999998,0.0821336791176748) (1,0.127324016899763)

\psline[linecolor=black] (0,0) (0.0125,-0.00512805085200476)
(0.025,-0.011270202591985) (0.0375,-0.0169199394032451)
(0.05,-0.0213713088653505) (0.0625,-0.0244916170787236)
(0.075,-0.0265328922200763) (0.0875,-0.0279780123239067)
(0.1,-0.0294176473635559) (0.1125,-0.031454412805345)
(0.125,-0.0346308717310956) (0.1375,-0.0393782553678828)
(0.15,-0.0459829974291655) (0.1625,-0.0545683960585031)
(0.175,-0.0650889283758839) (0.1875,-0.0773349466572658)
(0.2,-0.0909456820302712) (0.2125,-0.105428671243067)
(0.225,-0.120183904559304) (0.2375,-0.134531168149633)
(0.25,-0.147739222489647) (0.2625,-0.159055619235241)
(0.275,-0.167736112829294) (0.2875,-0.173072769698157)
(0.3,-0.174420017322945) (0.3125,-0.17121800771866)
(0.325,-0.163012794924184) (0.3375,-0.149472943997878)
(0.35,-0.13040229972688) (0.3625,-0.105748746793385)
(0.375,-0.0756088894983422) (0.3875,-0.0402286683214302) (0.4,0)
(0.4125,0.0445462965933687) (0.425,0.0927513115628269)
(0.4375,0.143841848989582) (0.45,0.19694731121141)
(0.4625,0.251118996238946) (0.475,0.305351455948178)
(0.4875,0.358605430159645) (0.5,0.409831859737499)
(0.5125,0.45799646444693) (0.525,0.502104361087603)
(0.5375,0.541224194381746) (0.55,0.574511257238628)
(0.5625,0.601229088323013) (0.575,0.620769053364164)
(0.5875,0.632667442301315) (0.6,0.636619647224958)
(0.6125,0.63249102609209) (0.625,0.62032410441212)
(0.6375,0.600341821474281) (0.65,0.57294658926839)
(0.6625,0.538715000968494) (0.675,0.498388101801183)
(0.6875,0.452857218189124) (0.7,0.403145431358823)
(0.712499999999999,0.350384879051035)
(0.724999999999999,0.295790173594924)
(0.737499999999999,0.240628336443251)
(0.749999999999999,0.186185768226636)
(0.762499999999999,0.133732899584857)
(0.774999999999999,0.0844873013481902)
(0.787499999999999,0.0395761731916195) (0.799999999999999,0)
(0.812499999999999,-0.0334064593056382)
(0.824999999999999,-0.0599992958934195)
(0.837499999999999,-0.0793604198137245)
(0.849999999999999,-0.0913153593286324)
(0.862499999999999,-0.0959471486139591)
(0.874999999999999,-0.0935991430851573)
(0.887499999999999,-0.0848651830940526)
(0.899999999999999,-0.070564599027648)
(0.912499999999999,-0.0516993388005176)
(0.924999999999999,-0.0293902789132972)
(0.937499999999999,-0.00478955391508862)
(0.949999999999999,0.0210344950055514)
(0.962499999999999,0.0472433924071538)
(0.974999999999999,0.0734095103425716)
(0.987499999999998,0.0997361033241759) (1,0.127320802599115)

\psline[linestyle=dotted] (0,0) (0.0125,0.0188332193789079)
(0.025,0.0367467008082227) (0.0375,0.052144502130601)
(0.05,0.0635848624521517) (0.0625,0.0699094371484411)
(0.075,0.0703534988114681) (0.0875,0.0646271671430882)
(0.1,0.0529597833033592) (0.1125,0.0361023137796075)
(0.125,0.0152859515560647) (0.1375,-0.00786137317350882)
(0.15,-0.0314353341420784) (0.1625,-0.0534039415743183)
(0.175,-0.0717735586867265) (0.1875,-0.0847592540804446)
(0.2,-0.0909456817667973) (0.2125,-0.089424901094047)
(0.225,-0.0798988672172992) (0.2375,-0.0627366692280825)
(0.25,-0.0389798143361772) (0.2625,-0.0102927257013981)
(0.275,0.0211401457690258) (0.2875,0.0527649201394266)
(0.3,0.0818469378324643) (0.3125,0.105684426269286)
(0.325,0.121831668673518) (0.3375,0.128314789702835)
(0.35,0.123823153196295) (0.3625,0.107860545503024)
(0.375,0.0808427417780894) (0.3875,0.044131573768485) (0.4,0)
(0.4125,-0.0484723843030913) (0.425,-0.0975688262839022)
(0.4375,-0.143160360394923) (0.45,-0.180972300825355)
(0.4625,-0.206874865031102) (0.475,-0.217178191983228)
(0.4875,-0.208911075183471) (0.5,-0.180063263231421)
(0.5125,-0.129773181964534) (0.525,-0.0584462853614232)
(0.5375,0.032206270743085) (0.55,0.139213622629205)
(0.5625,0.258475077219701) (0.575,0.384967269319714)
(0.5875,0.513016537855322) (0.6,0.63661977236758)
(0.6125,0.749793401480855) (0.625,0.84692799250337)
(0.6375,0.923125275784644) (0.65,0.974495358404431)
(0.6625,0.998394393035618) (0.675,0.993586851144206)
(0.6875,0.960321546537561) (0.7,0.900316316157109)
(0.712499999999999,0.81665238480812)
(0.724999999999999,0.713585487944896)
(0.737499999999999,0.596286375677887)
(0.749999999999999,0.47052798214593)
(0.762499999999999,0.342339992248736)
(0.774999999999999,0.217653535556404)
(0.787499999999999,0.101959153189267) (0.799999999999999,0)
(0.812499999999999,-0.0844804412139521)
(0.824999999999999,-0.14892084011753)
(0.837499999999999,-0.192044385895625)
(0.849999999999999,-0.213876355520873)
(0.862499999999999,-0.215678050777106)
(0.874999999999999,-0.199803936624571)
(0.887499999999999,-0.169493891186593)
(0.899999999999999,-0.128616616593877)
(0.912499999999999,-0.0813831819099677)
(0.924999999999999,-0.0320511887465925)
(0.937499999999999,0.0153599137390045)
(0.949999999999999,0.057323256376727)
(0.962499999999999,0.091012351133694)
(0.974999999999999,0.11444972871667)
(0.987499999999998,0.12658849635391) (1,0.127323954473516)


\psset{linewidth=0.3pt,linestyle=solid,linecolor=black}

\rput(0.5,-0.1){$0.5$} \rput(1,-0.1){$1$} \rput(1.15,0){$x$}

 \psline{->}(0,0)(1.1,0)

\psline(0.5,0)(0.5,-0.05)\psline(0.25,0)(0.25,-0.05)\psline(0.75,0)(0.75,-0.05)
\psline(1,0)(1,-0.05)\psline(0,0)(0,-0.05)

\psline(0,-0.5)(-0.05,-0.5) \psline(0,-0.25)(-0.05,-0.25)
\psline(0,-0.)(-0.05,-0.) \psline(0,0.5)(-0.05,0.5)
\psline(0,1)(-0.05,1)

\psline{->}(0,-0.51)(0,1.05)

\rput(-0.1,-0.5){$-0.5$} \rput(-0.1,-0.25){$-0.25$}
\rput(-0.1,0.5){$0.5$} \rput(-0.1,0){$0$}
\rput(-0.1,0.5){$0.5$} \rput(-0.1,1.0){$1$}

\psdots[dotsize=4pt] (0,0) (0.2,0) (0.4,0) (0.6,0) (0.8,0) (1,0)



\end{pspicture}

\caption{Sequence of smooth saturated models: $\hat y_0$ is a
polynomial of fifth degree \mbox{(- -)}, $\hat y_1,\ldots,\hat y_4$
\mbox{(---)} are supersaturated models. True model $m(x)$
\mbox{(...)} and design points are also shown.} \label{fig_1d}

\end{center}
\end{figure}
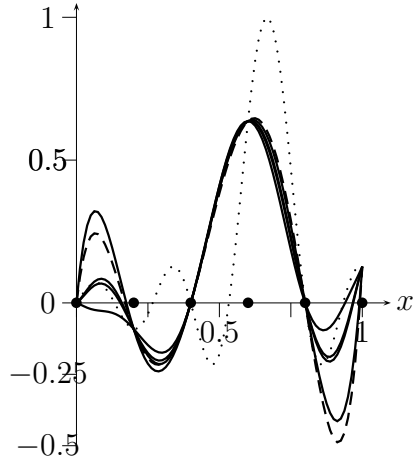

Suppose that the design $D_6$ is a uniform design in $[0,1]$, and
that the response vector $y$ contains the values of $m(x)$ at points
in $D_6$. The choice of a good saturated and supersaturated models
can be driven by algebraic methods. For the present case, an obvious
candidate is $h(x)=(1,x,\ldots,x^5)^T$. Call $\hat y_0$ to the
interpolator fitted solely with $h(x)$. Now a process of smoothing
is carried out by adding dummy points, one at a time. While adding
dummy points $h(x)$ remains unchanged. With only one dummy point, a
clear candidate for $g(x)$ is $g(x)=(x^6)$, while for $q$ dummy
points, $g(x)=(x^6,\ldots,x^{6+q-1})$ could be used. Call $\hat y_q$
to the smooth interpolator obtained by adding $q$ dummy points,
$q=1,\ldots,5$. The value of smoothness for $\hat y_q$ quickly drops
down so that a similar smoothness to that of a spline is achieved
with $\hat y_4$ (only four extra terms), see Table \ref{tab_d1}. The
progressive smoothing achieved with extra terms can be appreciated
graphically as well. Figure \ref{fig_1d} shows the interpolator and
smooth saturated models.

\begin{table}[h]
\begin{center}
\begin{tabular}{r|rrrrrr|r}Model&$\hat y_0$&$\hat y_1$&$\hat y_2$&$\hat
y_3$&$\hat y_4$&$\hat y_5$&Spline\\ \hline
$\Psi_2^*$&76.543&74.698&33.153& 33.020& 27.767& 27.745&26.744\\
\end{tabular}
\caption{Value of $\Psi^*_2$ for supersaturated models interpolating
$m(x)$ over $D_6$ of Section \ref{ex1}.}\label{tab_d1}
\end{center}
\end{table}

\begin{figure}[h!]  
\begin{center}
\psset{unit=48.3mm,linewidth=1.1pt}
 \begin{pspicture}(-0.2,-0.1)(1.1,1.1)
     \put(0,0){}

\psset{linecolor=black,linewidth=0.7pt}

\psline[linestyle=dashed] (0,0.299021252171618)
(0.1,0.291736696309992) (0.2,0.0497778422710041)
(0.3,0.0494620288663357) (0.4,0.00734785248409977)
(0.5,0.00716613076121405) (0.6,0.00183552179541712)
(0.7,0.00164563007375071) (0.8,0.00156114756741349)
(0.9,0.00155040195021991)   (1,0)

\psline[linestyle=dotted] (0,0.877306118317052)
(0.1,0.628785557608804) (0.2,0.179044293445035)
(0.3,0.0738246111800731) (0.4,0.028839371969689)
(0.5,0.0202613789033006) (0.6,0.0149953994505871)
(0.7,0.0131278914667052) (0.8,0.0115817655489575)
(0.9,0.0054657744670298) (1,0.00529732768480023)

\psline (0,0.466642954124052)   (0.1,0.392528918561805)
(0.2,0.00462818325598107)   (0.3,0.00217564531807273)
(0.4,0.00215584109562133)   (0.5,0.00197114762889543)
(0.6,0.00186462389947655)   (0.7,0.0018431086433937)
(0.8,0.00175237160673308)   (0.9,0) (1,0)

\psline (0,0.00232150091999855) (0.1,0.00156232413797553)
(0.2,0.0014922655655914)    (0.3,0.00112352577287441)
(0.4,0.000664700811392804)  (0.5,0) (0.6,0) (0.7,0) (0.8,0) (0.9,0)
(1,0)


\psset{linewidth=0.3pt,linestyle=solid,linecolor=black}

\rput(1.15,0){$q$}

\rput(-0.25,0.8){$R(q)$}

 \psline{-}(0,1)(0,0)(1,0)(1,1)(0,1)

\psline(0,0.01)(0,0) \psline(0.1,0.01)(0.1,0)
\psline(0.2,0.01)(0.2,0) \psline(0.3,0.01)(0.3,0)
\psline(0.4,0.01)(0.4,0) \psline(0.5,0.01)(0.5,0)
\psline(0.6,0.01)(0.6,0) \psline(0.7,0.01)(0.7,0)
\psline(0.8,0.01)(0.8,0) \psline(0.9,0.01)(0.9,0)
\psline(1,0.01)(1,0)

\psline(0,0.99)(0,1) \psline(0.1,0.99)(0.1,1)
\psline(0.2,0.99)(0.2,1) \psline(0.3,0.99)(0.3,1)
\psline(0.4,0.99)(0.4,1) \psline(0.5,0.99)(0.5,1)
\psline(0.6,0.99)(0.6,1) \psline(0.7,0.99)(0.7,1)
\psline(0.8,0.99)(0.8,1) \psline(0.9,0.99)(0.9,1)
\psline(1,0.99)(1,1)

\psline(0.01,0)(0,0)
\psline(0.01,0.301029995663981)(0,0.301029995663981)
\psline(0.01,0.477121254719662)(0,0.477121254719662)
\psline(0.01,0.602059991327962)(0,0.602059991327962)
\psline(0.01,0.698970004336019)(0,0.698970004336019)
\psline(0.01,0.778151250383644)(0,0.778151250383644)
\psline(0.01,0.845098040014257)(0,0.845098040014257)
\psline(0.01,0.903089986991944)(0,0.903089986991944)
\psline(0.01,0.954242509439325)(0,0.954242509439325)
\psline(0.01,1)(0,1)

\psline(0.99,0)(1,0)
\psline(0.99,0.301029995663981)(1,0.301029995663981)
\psline(0.99,0.477121254719662)(1,0.477121254719662)
\psline(0.99,0.602059991327962)(1,0.602059991327962)
\psline(0.99,0.698970004336019)(1,0.698970004336019)
\psline(0.99,0.778151250383644)(1,0.778151250383644)
\psline(0.99,0.845098040014257)(1,0.845098040014257)
\psline(0.99,0.903089986991944)(1,0.903089986991944)
\psline(0.99,0.954242509439325)(1,0.954242509439325)
\psline(0.99,1)(1,1)


\rput(0,-0.1){$0$} \rput(0.2,-0.1){$2$} \rput(0.4,-0.1){$4$}
 \rput(0.6,-0.1){$6$} \rput(0.8,-0.1){$8$}
\rput(1,-0.1){$10$}

\rput(-0.05,0){$1$} \rput(-0.05,0.301029995663981){$2$}
\rput(-0.05,0.477121254719662){$3$}
\rput(-0.05,0.602059991327962){$4$}
\rput(-0.05,0.698970004336019){$5$}
 \rput(-0.05,1){$10$}

\end{pspicture}

\caption{Logarithm of smoothness ratio
$R(q)=\sqrt{\Psi_2^*(q)/\Psi_2^*(\mbox{sp})}$ against number of
smoothing terms added $q$: sample sizes $n=5,10,15$ \mbox{(-
-,...,---)}. The line for $n=20$ is indistinguishable from
$R(q)=1$.} \label{fig_1d2}

\end{center}
\end{figure}
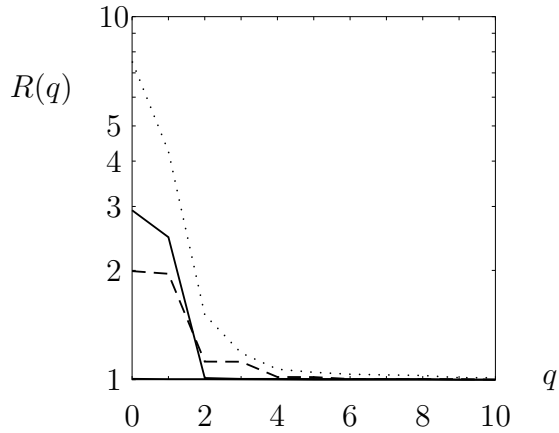

A comparison between the smooth supersaturated method and cubic
splines, which are optimally smooth, is carried out as follows.
First, for a uniform design $D_n$ on $[0,1]$, a saturated model
$\hat y_0$ is fitted to the values of $m(x)$ at the design points.
Call $\Psi_2^*(0)$ the value of smoothness for $\hat y_0$. Then,
using extra $q$ basis terms, a smooth supersaturated model $\hat
y_q$ is fitted. Call $\Psi_2^*(q)$ the corresponding value of
smoothness. Additionally, a spline is fitted to the same data and
call $\Psi_2^*(\mbox{sp})$ its smoothness value. The important
feature is that the $\Psi_2^*(0),\Psi_2^*(1),\ldots$ form a
decreasing sequence which converges surprisingly quick to
$\Psi_2^*(\mbox{sp})$. This behavior can be quantified by plotting
the ratio $\sqrt{\Psi_2^*(q)/\Psi_2^*(\mbox{sp})}$ against the
number of terms added to smooth the model. Figure \ref{fig_1d2}
shows such comparison when $D_n$ are uniform designs of size
$n=5,10,15,20$.

\subsection{A two dimensional example: alternative to thin-plate splines?}

\label{ex2} The objective of this example is to compare the
performance of smooth supersaturated interpolators against thin
plate splines, but there is also interest to make comparisons
against a kriging interpolator. Initially, interpolators of the
three kinds above are constructed for a known function at given
design points and then predictions over new design points are used
to compare the performance of the interpolators. The known function
is $m(x_1,x_2)$, which is constructed as
$m(x_1,x_2)=p(4x_1-2,4x_2-2)$, where
$p(x_1,x_2)$ is the \texttt{peaks} function from
MATLAB\textregistered. The objective of scaling and shifting
$p(x_1,x_2)$ is to include interesting features into the smoothing
region $\mathcal X=[0,1]^2$.

In order to allow a good covering of the design region $\mathcal X$
without an excessive number of points, we use Sobol's space filling
design $D_{24}$ and $h(x)$ to be the good saturated model of Example
\ref{ex_sobol}. The response vector $y$ contains the values of
$p(x_1,x_2)$ at points in $D_{24}$.

A smooth supersaturated model was then fitted to this data using the
$91$ terms of a good supersaturated complete model of degree twelve
in $x_1,x_2$. Call this model $\hat y$. A thin plate spline
interpolator model was also fitted to the same data, which we refer
to as $\hat y_{sp}$. A kriging interpolator, $\hat y_{kr}$, was also
fitted using the model
\begin{equation}\label{ec_krig}Y(x)=\beta+Z(x),\end{equation} where
$Z(x)$ is a stochastic process with exponential covariance
structure, i.e.
$\mbox{Cov}(Z(r),Z(s))=\exp(\sum_{i=1}^2\theta_i|r_i-s_i|^{p_i})$.

\begin{figure}[h!]  
\begin{center}
\psset{unit=3.0mm,linewidth=0.85pt}
 \begin{pspicture}(-10,-12)(10,11)
     \put(0,0){}

\psset{linecolor=black}


\psdots[dotsize=3pt]
(2.52323973,2.627844873)
(2.24275589,3.775017232)
(4.54063436,5.678613789)
(2.15297424,2.398787689)
(2.64242987,2.66753671)
(-7.26006482,-8.325055083)
(-3.38843814,-3.50859752)
(-1.11940195,-1.282328189)
(3.59206185,3.98052527)
(-3.90553369,-5.397169546)
(0.95921724,0.89090631)
(-2.85985271,-2.590265361)
(1.64605616,1.49826681)
(1.10099795,0.9605137703)
(1.45646942,1.554615297)
(-0.82960019,-0.4868283817)
(2.17538763,2.575943271)
(2.1689318,3.805403597)
(4.14700148,4.871757466)
(2.52587097,2.742507468)
(1.9207745,1.528787946)
(-3.39131079,-3.659410646)
(-2.14136371,-2.31140841)
(-0.48464051,-0.5578475178)
(3.26995646,3.519012498)
(2.23627324,3.411621777)
(4.55431819,6.034284451)
(2.12158154,2.273504707)
(0.61617251,-0.09690062294)
(-4.72028519,-5.076233684)
(-1.27230106,-1.670028519)
(-1.10138865,-1.192354622)
(5.26245867,5.116926356)
(0.27953291,-0.01416654384)
(1.3099938,1.183009218)
(-1.9014544,-1.588246607)
(2.1642379,2.807541421)
(-0.04812015,-0.1765101336)
(-2.06885211,-2.855657844)
(-2.58611967,-3.023353796)
(5.29269757,5.73463237)
(-2.28705212,-2.780515147)
(-0.17841159,-0.6573419382)
(-5.14108371,-5.760788934)
(2.11448315,2.894762614)
(-0.98709301,-1.448205123)
(-0.72193718,-0.9733922831)
(1.12472685,0.8453479166)
(2.97271084,3.193242054)
(2.16329264,3.33854127)
(5.57627079,6.543790194)
(3.17386288,3.488591733)
(1.77999476,1.82223938)
(-6.68535107,-7.02422391)
(-2.96345818,-2.881888781)
(-0.14920712,0.4795575467)
(1.98470821,1.822306389)
(2.09459688,3.546882613)
(3.63238471,4.66178512)
(1.06029187,1.10723608)
(1.09463844,-0.8616599753)
(-0.57717082,-0.6380465386)
(0.01431977,-0.04566742967)
(0.40581339,0.3393841142)
(3.96722772,4.00240948)
(-0.43469921,-0.5496043174)
(1.09226509,1.123148026)
(0.78426102,0.7448846056)
(2.36685182,2.414442117)
(0.55538898,0.5653902311)
(0.0370783,-0.05447153561)
(-0.88270572,0.03217984645)
(1.78248217,1.538079273)
(1.98582349,3.039312256)
(2.98265986,3.701772003)
(1.00363086,1.448785053)
(1.18098198,0.8684140376)
(-6.82192885,-8.292449657)
(-1.38664247,-1.487598819)
(0.53743671,0.5917251247)
(4.19581582,4.107423576)
(1.45522352,2.289612134)
(2.19687137,2.282535172)
(-2.99459391,-3.044032638)
(2.14425283,2.654719085)
(-0.23889504,-0.7385731162)
(-0.67233904,-1.044446335)
(-1.58142314,-1.961234691)
(4.82835045,5.088436094)
(-2.09031973,-3.019964179)
(0.4670119,0.08848158648)
(-0.59116056,-0.2572211669)
(1.12418248,0.6817004083)
(1.60576734,1.869121677)
(2.75586901,3.662506523)
(-1.20333293,-0.06039267212)
(1.40057739,1.207182974)
(1.56755193,2.458275517)
(2.65286827,3.303710171)
(1.56242477,1.788463848)
(3.45167429,3.393995204)
(-3.96851593,-3.92975832)
(-3.40630217,-3.903484432)
(-2.11814128,-2.548195214)
(4.37215696,4.972606857)
(-3.82240929,-4.691973254)
(-0.20215006,-0.6534973466)
(0.11246408,0.5090169776)
(1.6416825,1.403772238)
(0.98806439,0.8677802806)
(1.63648262,2.129790571)
(-1.78098887,-0.8653304294)
(1.77136432,2.042280526)
(1.78321232,2.826105562)
(3.01896587,3.594842016)
(3.04994038,3.28268103)
(2.08984374,2.110867934)
(-5.27896951,-5.363115189)
(-2.9692101,-3.0098337)
(0.05970666,0.2542782036)
(2.5952268,2.629523333)
(2.24625639,3.721020467)
(4.26163964,5.485603547)
(0.68370111,0.7835269699)
(0.62774397,-0.255734196)
(-5.04757319,-6.063753316)
(-0.49143293,-0.8375556905)
(-1.14388645,-1.229149434)
(5.34230311,5.322537075)
(1.32221823,1.828409882)
(2.30428534,2.645406587)
(-0.70402922,-0.6542220126)
(2.30333804,2.652432806)
(0.51390755,0.3414910559)
(-0.92261659,-1.440854112)
(-1.18036381,0.01509667654)
(1.22539813,0.8693858149)
(1.47536474,2.313617471)
(2.17518847,2.698522859)
(2.21182759,2.405133296)
(1.03777478,0.1147154057)
(-2.98060636,-3.161531568)
(-1.3078123,-1.5755806)
(1.6044876,1.572272665)
(2.98543681,3.093036599)
(-0.21972459,0.4480248846)
(1.00104176,1.00087844)
(-1.87334819,-1.704649265)
(1.97733455,2.026225758)
(0.52213082,0.1835133029)
(0.45219493,0.3323405695)
(-1.85767786,-2.055281271)
(5.48108008,5.479173874)
(-0.72850048,-1.291587615)
(0.73323145,0.4238666351)
(-6.39050659,-7.166426858)
(1.89810675,2.380456444)
(0.11293003,-0.0209519814)
(0.38166725,0.2983573339)
(0.53725579,0.2447598666)
(3.22400994,3.444242221)
(2.2175756,3.420299184)
(5.20141769,6.547412916)
(2.34367278,2.684669014)
(2.81386607,2.819239168)
(-6.16181009,-6.380083525)
(-3.50408949,-3.678124917)
(-1.11668036,-0.6356074522)
(2.5106742,2.607148473)
(2.072021,3.051956799)
(3.52770209,4.551215978)
(2.83359987,3.120621839)
(1.05124192,0.9189461476)
(-6.43638768,-6.879249271)
(-2.14243387,-2.216864258)
(0.58006022,0.679740554)
(4.18198604,4.068236195)
(0.80257096,1.173365957)
(1.43080643,1.276772403)
(-3.87445557,-4.090677019)
(2.22317664,3.220111655)
(-1.55565357,-2.00541582)
(-1.75178112,-2.156570941)
(-0.46893017,-0.797884014)
(4.09725399,4.326802657)
(-1.95431284,-2.91664476)
(1.17926363,1.164840228)
(-3.49104934,-3.363747721)
(1.20592497,0.704835473)
(1.84374908,2.455127189)
(2.78733252,3.396501974)
(-0.04424975,0.351930509)
(2.13602064,2.22466658)
(2.14567018,3.651434846)
(4.05626988,5.038097601)
(1.51891107,1.62758707)
(3.09778647,2.802560631)
(-2.30050073,-2.5720765)
(-2.47152154,-2.937331019)
(-3.00323512,-3.216416284)
(5.98194215,6.355235231)
(-0.60151211,-0.6910182443)
(0.36474687,0.2162232382)
(-4.02865163,-4.020300148)
(1.8796919,2.067419937)
(0.49508949,0.1012611267)
(0.54152808,0.3609517788)
(0.63916858,0.5286039602)
(2.54918798,2.806002971)
(2.23986497,3.790338765)
(5.05132827,5.860377931)
(2.34534128,2.78486689)
(1.76346353,1.679321266)
(-7.36784937,-8.50803814)
(-2.62847907,-2.558203806)
(-1.20868745,0.005634908937)
(1.1919357,0.8383188922)
(1.41743919,2.244289325)
(2.14479528,2.67056069)
(2.17482662,2.358692684)
(0.93314811,-0.02159357759)
(-3.0792802,-3.250950615)
(-1.25938216,-1.556634673)
(-0.88888509,-1.093987302)
(4.907575,4.870010287)
(-0.55955175,-1.214166877)
(0.93855237,0.7066807478)
(1.65006668,1.617425486)
(2.87934896,2.984940031)
(0.9603955,0.959494669)
(0.13036653,0.05569085578)
(-3.46930047,-3.803050039)
(6.09977347,6.866469785)
(-0.2422266,-0.2219034312)
(0.28697544,0.2934501506)
(-2.30029013,-2.1319896)
(1.93484955,1.987697577)
(0.55606653,0.2000372253)
(0.5359885,0.4154171654)
(0.90178726,0.6447993943)
(2.81673682,3.043507787)
(2.22815853,3.613775319)
(5.19910118,6.227302556)
(2.77944469,3.168775554)
(2.09061641,2.100544528)
(-7.06087896,-7.713981034)
(-3.10950464,-3.049377607)
(-1.43174268,-0.1186507342)
(1.12021582,0.8285592264)
(1.24295968,1.928751522)
(2.11633961,2.597750855)
(1.73861691,1.874407033)
(0.63159648,-0.5793155228)
(-3.2168286,-3.488641609)
(-0.90939335,-1.347014267)
(-0.10977551,-0.2100504513)
(4.41331116,4.380362597)
(-0.29794396,-0.6533840644)
(1.03426869,0.928592673)
(1.52415752,1.515413691)
(2.52035023,2.569650651)
(0.72824142,0.7219624529)
(0.22837756,0.2271618016)
(-0.81547066,-0.7670942464)
(3.14436077,3.404113425)
(2.20064607,3.312780234)
(4.24743477,5.652752942)
(2.66632561,2.822547522)
(0.73695567,0.3319228698)
(-5.30518711,-5.550454626)
(-1.6923456,-1.964540901)
(1.10387502,1.266793857)
(3.80736306,3.728177333)
(1.38219955,2.3246242)
(1.9627278,1.903105687)
(-3.9203569,-4.164955146)
(2.14602757,2.829551131)
(-0.69610013,-1.225160316)
(-0.8514659,-1.186137389)
(-0.61365041,-0.860518297)
(3.6881039,3.996892945)
(-3.04235382,-4.356528075)
(1.38556314,1.429786818)
(-3.86370161,-3.647906466)
(1.46217441,1.221821987)
(1.50413751,1.656455391)
(2.02801091,2.251437477)
(-0.56120912,0.2519800511)
(1.7637157,1.654029425)
(1.93184414,3.207380407)
(3.34765388,4.241680718)
(1.43229512,1.585759065)
(3.41823295,3.315799803)
(-2.91496033,-2.985417119)
(-3.02685693,-3.564282479)
(-1.17174013,-0.03378114412)
(1.40109215,1.085272572)
(1.71261047,2.49549392)
(2.37573998,2.851758483)
(2.0634495,2.279923634)
(1.563824,0.6475991465)
(-2.35369378,-2.668855848)
(-1.34414824,-1.560226485)
(1.16914684,1.174388977)
(3.47336166,3.535906971)
(-0.12703585,0.2509355922)
(1.07412577,1.066294965)
(-0.22838231,-0.1027430713)
(2.11556414,2.099845296)
(0.51083899,0.3445741847)
(0.39210938,0.3659899104)
(-1.06780888,-1.271819126)
(5.03681368,4.988241639)
(-0.54139645,-1.195019001)
(0.9247984,0.6723424798)
(-7.20402909,-8.623521008)
(1.95181125,2.622155285)
(-0.15853056,-0.1453259876)
(0.29216823,0.3059731711)
(0.8837675,0.6067975927)
(2.89380054,3.117177005)
(2.22523401,3.572010479)
(5.21501787,6.305361623)
(2.77519949,3.1539826)
(2.20831174,2.228898772)
(-6.93016555,-7.48491991)
(-3.20555301,-3.170514927)
(-2.03268228,-2.450275547)
(4.95629621,5.322389826)
(-2.44807464,-3.237002844)
(0.12779768,-0.3412014784)
(0.38562949,0.8149616227)
(1.22399269,0.8818173728)
(1.42674895,1.483154852)
(2.49862405,3.423971083)
(-1.36554322,-0.8696203184)
(2.07340117,2.528036198)
(2.08519481,3.611744321)
(3.76509892,4.440151307)
(2.95219141,3.132865846)
(1.73960311,1.545226576)
(-4.2514383,-4.397168236)
(-2.40381216,-2.500058294)
(-0.37589808,0.2601884336)
(2.11909164,1.960560282)
(2.14301287,3.524574349)
(3.58560225,4.595450024)
(0.60015346,0.8923374744)
(0.87100564,0.3452980641)
(-6.03375927,-7.336112702)
(-0.78187555,-0.9950330956)
(-1.46598292,-1.508230216)
(5.54856887,5.444056177)
(0.7080103,0.7306525506)
(1.66201802,1.729688825)
(-1.01329708,-0.8167998405)
(2.24936243,2.72637919)
(0.62699372,0.5395838301)
(-1.64610264,-2.394344905)
(-1.19499376,-0.3867616969)
(2.02422143,1.935663323)
(1.97362829,2.822494706)
(3.04753669,3.774621486)
(2.30466875,2.742160312)
(1.28373062,1.130384479)
(-6.94225182,-7.838232297)
(-2.13885732,-2.160217442)
(-0.4465487,-0.5341982222)
(4.8665685,4.808728767)
(1.47418406,2.152724529)
(2.42469304,2.723648342)
(-1.47274458,-1.411958491)
(2.21623404,2.564102485)
(0.22423782,-0.1076444589)
(-0.64125534,-1.060970455)
(-0.86082722,-1.201650164)
(4.46296239,4.65962181)
(-1.76444426,-2.734225271)
(0.88724925,0.714979282)
(-2.28628454,-2.119853522)
(1.10172451,0.5454298023)
(1.82145017,2.390676423)
(2.95934036,3.753654574)
(-1.78813005,-0.590593932)
(1.47776147,1.49724976)
(1.50480621,2.220871959)
(2.58018842,3.077863593)
(2.52087261,2.83326749)
(2.67713572,2.685287167)
(-5.32678017,-5.40758496)
(-3.30223072,-3.471384014)
(-2.50077355,-2.717537011)
(5.73350813,5.89616831)
(-0.9443643,-1.302302916)
(0.42889737,0.119833805)
(-5.33465555,-5.68840204)
(1.91102143,2.300880059)
(0.17473486,-0.1679739927)
(0.31780025,0.1320111864)
(1.54507845,1.460699385)
(2.96070546,3.087835346)
(2.00588705,2.804415924)
(6.02498485,6.866401057)
(3.55370162,3.587146749)
(1.25459667,1.283636362)
(-5.94159018,-5.968219472)
(-2.57675229,-2.552930586)
(-0.70725409,0.2853601014)
(1.54817538,1.223939641)
(1.82559135,3.005705582)
(2.87357205,3.665532919)
(1.5725483,1.73724136)
(1.01689214,-0.388388298)
(-1.84515452,-2.02727835)
(-0.71114642,-0.9536024185)
(0.18527468,-0.09356192674)
(3.82297604,4.000496518)
(-1.40049294,-1.931558864)
(1.39217948,1.507961046)
(0.38111163,0.3087209175)
(2.44278247,2.670960356)
(0.85619119,0.8661780252)
(-0.70509538,-1.122907585)
(-1.75980957,-2.055157352)
(3.92951237,4.477432334)
(-4.18107982,-5.35993361)
(0.15034967,-0.1629446001)
(-1.03281783,-0.7024925994)
(1.71800401,1.521677183)
(0.92600491,0.7717110375)
(1.36402787,1.612873822)
(0.01353983,0.2325804977)
(2.27162216,2.502023695)
(2.20086228,3.785834446)
(4.39155229,5.284734995)
(1.77298617,1.918359383)
(2.6565748,2.168762546)
(-2.30255361,-2.684231906)
(-2.0941405,-2.437421204)
(0.09726103,0.04224529331)
(2.97599158,3.133484902)
(2.25946838,3.627549744)
(4.63651496,5.985249068)
(1.08301734,1.141959485)
(0.53594382,-0.5299747744)
(-4.47306995,-5.227006571)
(-0.59928659,-1.04275014)
(-0.19665211,-0.2291794806)
(4.67575928,4.539219729)
(0.43450719,0.3281917835)
(1.30761291,1.159564811)
(-3.10305281,-3.008842311)
(2.19891239,3.109755718)
(-1.18558252,-1.512254768)
(-2.05992878,-2.642344609)
(-1.7432366,-0.2318170899)
(0.94919447,0.7103726036)
(0.79234073,1.190914301)
(1.26675644,1.502772043)
(2.95962939,3.073438355)
(1.00738418,0.5947124135)
(-4.40107862,-4.47960128)
(-1.89552551,-2.084304536)
(1.08650422,1.260910974)
(3.75585834,3.695139497)
(0.58551404,1.129982314)
(1.17849873,1.004323784)
(1.19680744,1.493660981)
(1.971242,1.835878238)
(0.73386341,0.5793319356)
(1.13166878,1.526777131)
(-2.47261434,-2.634166474)
(5.81027241,5.920189663)
(-0.65889594,-0.9597335366)
(0.5922843,0.339429715)
(-5.37375051,-5.647205541)
(1.85308114,2.156494891)
(0.43216793,0.1406061925)
(0.57729743,0.4171053288)
(0.03201539,0.5057641094)
(2.12843385,2.035336088)
(2.15164214,3.655027778)
(3.87072178,4.951162809)
(1.66079118,1.645949093)
(2.91980096,3.020468087)
(-7.2809754,-8.788729565)
(-3.51956976,-3.877127759)
(-1.58376738,-1.504461719)
(3.14375891,3.554089342)
(-5.11654085,-6.805730899)
(0.43408856,0.3284452833)
(-2.13645461,-1.963230732)
(1.94467079,1.989398308)
(0.55379156,0.2063649355)
(0.52654032,0.4123572713)
(-0.84184789,-0.1675363139)
(1.9236045,2.088290801)
(2.00210859,3.335135634)
(3.58592072,4.401116189)
(2.37691526,2.613883424)

\psset{linewidth=0.3pt,linestyle=solid,linecolor=black}

\psline(-9,-9)(-9,-8.85)    \psline(-9,7)(-9,6.85)  \psline(-9,-9)(-8.85,-9)    \psline(7,-9)(6.85,-9)
\psline(-7,-9)(-7,-8.85)    \psline(-7,7)(-7,6.85)  \psline(-9,-7)(-8.85,-7)    \psline(7,-7)(6.85,-7)
\psline(-5,-9)(-5,-8.85)    \psline(-5,7)(-5,6.85)  \psline(-9,-5)(-8.85,-5)    \psline(7,-5)(6.85,-5)
\psline(-3,-9)(-3,-8.85)    \psline(-3,7)(-3,6.85)  \psline(-9,-3)(-8.85,-3)    \psline(7,-3)(6.85,-3)
\psline(-1,-9)(-1,-8.85)    \psline(-1,7)(-1,6.85)  \psline(-9,-1)(-8.85,-1)    \psline(7,-1)(6.85,-1)
\psline(1,-9)(1,-8.85)  \psline(1,7)(1,6.85)    \psline(-9,1)(-8.85,1)  \psline(7,1)(6.85,1)
\psline(3,-9)(3,-8.85)  \psline(3,7)(3,6.85)    \psline(-9,3)(-8.85,3)  \psline(7,3)(6.85,3)
\psline(5,-9)(5,-8.85)  \psline(5,7)(5,6.85)    \psline(-9,5)(-8.85,5)  \psline(7,5)(6.85,5)
\psline(7,-9)(7,-8.85)  \psline(7,7)(7,6.85)    \psline(-9,7)(-8.85,7)  \psline(7,7)(6.85,7)

\psline(-9,-9)(-9,7)(7,7)(7,-9)(-9,-9)

\psline[linestyle=dashed](-9,-9)(7,7)

\rput(-10.1,-9){$-9$}   \rput(-9,-10.1){$-9$}
\rput(-10.1,-5){$-5$}   \rput(-5,-10.1){$-5$}
\rput(-10.1,-1){$-1$}   \rput(-1,-10.1){$-1$}
\rput(-10.1,3){$3$} \rput(3,-10.1){$3$}
\rput(-10.1,7){$7$} \rput(7,-10.1){$7$}

\rput(0.5,-11.5){$\hat y_{sp}$}
\rput(-11.5,0.5){$\hat y$}
\rput(-0.5,9.5){(a)}

\end{pspicture}
 \begin{pspicture}(-12.5,-12)(7.5,11)
     \put(0,0){}

\psset{linecolor=black}


\psdots[dotsize=3pt]
(2.0012569,2.627844873)
(2.8886783,3.775017232)
(5.7410256,5.678613789)
(2.4871005,2.398787689)
(2.5932865,2.66753671)
(-6.9854903,-8.325055083)
(-2.2996209,-3.50859752)
(-1.0764838,-1.282328189)
(3.8349737,3.98052527)
(-2.5395087,-5.397169546)
(0.89597095,0.89090631)
(-2.944895,-2.590265361)
(1.740274,1.49826681)
(1.0262214,0.9605137703)
(1.4107358,1.554615297)
(0.10796947,-0.4868283817)
(2.4075129,2.575943271)
(1.4381808,3.805403597)
(5.1357092,4.871757466)
(2.8276385,2.742507468)
(1.3680008,1.528787946)
(-4.018641,-3.659410646)
(-1.9747196,-2.31140841)
(-0.61911417,-0.5578475178)
(3.126458,3.519012498)
(3.4527071,3.411621777)
(4.9327735,6.034284451)
(2.170026,2.273504707)
(0.33401224,-0.09690062294)
(-4.431828,-5.076233684)
(-1.8471221,-1.670028519)
(-1.4490205,-1.192354622)
(5.4390637,5.116926356)
(-0.10188645,-0.01416654384)
(1.1235751,1.183009218)
(-0.63379064,-1.588246607)
(2.838779,2.807541421)
(-0.50423338,-0.1765101336)
(-1.0371142,-2.855657844)
(-2.0470566,-3.023353796)
(6.2225283,5.73463237)
(-2.1587938,-2.780515147)
(-0.68427635,-0.6573419382)
(-4.3928682,-5.760788934)
(2.8851664,2.894762614)
(-1.5924955,-1.448205123)
(-0.81423251,-0.9733922831)
(0.97753338,0.8453479166)
(2.7337147,3.193242054)
(2.2983475,3.33854127)
(6.325626,6.543790194)
(3.478848,3.488591733)
(1.8233858,1.82223938)
(-6.7115481,-7.02422391)
(-2.5654907,-2.881888781)
(0.61262621,0.4795575467)
(1.2636849,1.822306389)
(2.8934176,3.546882613)
(4.7541804,4.66178512)
(1.129253,1.10723608)
(0.09096172,-0.8616599753)
(-0.62670283,-0.6380465386)
(-0.056593375,-0.04566742967)
(0.33936334,0.3393841142)
(3.9628406,4.00240948)
(-0.28700368,-0.5496043174)
(1.1508464,1.123148026)
(0.7825133,0.7448846056)
(2.3856695,2.414442117)
(0.56569284,0.5653902311)
(0.11195611,-0.05447153561)
(0.37424906,0.03217984645)
(1.0980539,1.538079273)
(2.9080271,3.039312256)
(3.623752,3.701772003)
(1.0893129,1.448785053)
(1.0273728,0.8684140376)
(-5.6327965,-8.292449657)
(-1.8027144,-1.487598819)
(0.40263948,0.5917251247)
(3.9475031,4.107423576)
(0.44315649,2.289612134)
(2.0236926,2.282535172)
(-2.4838679,-3.044032638)
(2.8933691,2.654719085)
(-0.86323895,-0.7385731162)
(-0.53573807,-1.044446335)
(-1.7680368,-1.961234691)
(5.5103542,5.088436094)
(-1.8241698,-3.019964179)
(0.12223157,0.08848158648)
(-0.77291551,-0.2572211669)
(0.90535007,0.6817004083)
(1.8955338,1.869121677)
(2.5778103,3.662506523)
(0.90485626,-0.06039267212)
(0.960006,1.207182974)
(1.7069039,2.458275517)
(3.3714141,3.303710171)
(1.7076466,1.788463848)
(3.3175353,3.393995204)
(-4.4917389,-3.92975832)
(-1.7493629,-3.903484432)
(-1.6326793,-2.548195214)
(5.2806943,4.972606857)
(-3.2671906,-4.691973254)
(-0.7170761,-0.6534973466)
(0.16224187,0.5090169776)
(1.5875214,1.403772238)
(0.87890215,0.8677802806)
(1.4150614,2.129790571)
(0.33757548,-0.8653304294)
(1.9643645,2.042280526)
(1.1523178,2.826105562)
(3.588342,3.594842016)
(3.3042843,3.28268103)
(2.0128007,2.110867934)
(-5.7108602,-5.363115189)
(-2.3394948,-3.0098337)
(0.1279055,0.2542782036)
(2.0078912,2.629523333)
(3.3840485,3.721020467)
(5.2440234,5.485603547)
(0.46330257,0.7835269699)
(0.30049065,-0.255734196)
(-4.1814729,-6.063753316)
(-1.0022602,-0.8375556905)
(-1.5504586,-1.229149434)
(4.9087887,5.322537075)
(0.86112122,1.828409882)
(2.5841523,2.645406587)
(-0.36166486,-0.6542220126)
(2.6736939,2.652432806)
(0.2475432,0.3414910559)
(-0.46987981,-1.440854112)
(0.92549805,0.01509667654)
(0.57135954,0.8693858149)
(2.0562654,2.313617471)
(2.7624473,2.698522859)
(2.4816418,2.405133296)
(0.43131101,0.1147154057)
(-3.2126076,-3.161531568)
(-1.6538905,-1.5755806)
(1.7194578,1.572272665)
(2.8639502,3.093036599)
(0.47717092,0.4480248846)
(1.0012107,1.00087844)
(-1.7968161,-1.704649265)
(2.2940937,2.026225758)
(0.20088383,0.1835133029)
(0.44220771,0.3323405695)
(-2.0312261,-2.055281271)
(5.9712344,5.479173874)
(-0.79603212,-1.291587615)
(0.42824397,0.4238666351)
(-6.0415946,-7.166426858)
(2.0263482,2.380456444)
(-0.065205956,-0.0209519814)
(0.31950092,0.2983573339)
(0.20946604,0.2447598666)
(2.9193888,3.444242221)
(3.0455326,3.420299184)
(5.8800187,6.547412916)
(2.7098812,2.684669014)
(2.8288814,2.819239168)
(-6.4017711,-6.380083525)
(-2.1640854,-3.678124917)
(-0.35032439,-0.6356074522)
(2.2311565,2.607148473)
(3.0981853,3.051956799)
(3.8981529,4.551215978)
(2.9899725,3.120621839)
(1.0069495,0.9189461476)
(-6.0717126,-6.879249271)
(-2.4223178,-2.216864258)
(0.53632363,0.679740554)
(4.2366187,4.068236195)
(0.078787269,1.173365957)
(1.0890426,1.276772403)
(-2.5265544,-4.090677019)
(3.3412091,3.220111655)
(-2.3414216,-2.00541582)
(-1.3673199,-2.156570941)
(-0.85559684,-0.797884014)
(4.2835343,4.326802657)
(-1.2469083,-2.91664476)
(1.2555613,1.164840228)
(-3.9407549,-3.363747721)
(0.92737398,0.704835473)
(2.5060583,2.455127189)
(3.0425862,3.396501974)
(0.63120897,0.351930509)
(1.7577602,2.22466658)
(2.2885022,3.651434846)
(5.2691069,5.038097601)
(1.6329014,1.62758707)
(2.38031,2.802560631)
(-3.062849,-2.5720765)
(-1.6434798,-2.937331019)
(-2.2882855,-3.216416284)
(6.3654751,6.355235231)
(-0.52928048,-0.6910182443)
(0.21575923,0.2162232382)
(-3.8754355,-4.020300148)
(2.2787006,2.067419937)
(0.10169071,0.1012611267)
(0.45188718,0.3609517788)
(0.71495383,0.5286039602)
(2.3802919,2.806002971)
(2.0253904,3.790338765)
(6.1907095,5.860377931)
(2.6886742,2.78486689)
(1.7259324,1.679321266)
(-6.8251682,-8.50803814)
(-2.5249824,-2.558203806)
(0.97060019,0.005634908937)
(0.55404317,0.8383188922)
(1.9595475,2.244289325)
(2.7374878,2.67056069)
(2.4217211,2.358692684)
(0.36320368,-0.02159357759)
(-3.2436245,-3.250950615)
(-1.6509987,-1.556634673)
(-1.2459505,-1.093987302)
(5.2807348,4.870010287)
(-0.7276084,-1.214166877)
(0.70929043,0.7066807478)
(1.7637541,1.617425486)
(2.7581431,2.984940031)
(0.95904727,0.959494669)
(-0.021215679,0.05569085578)
(-2.0843713,-3.803050039)
(6.2223266,6.866469785)
(-0.21409512,-0.2219034312)
(0.2954213,0.2934501506)
(-2.224855,-2.1319896)
(2.2651523,1.987697577)
(0.2213721,0.2000372253)
(0.50642727,0.4154171654)
(0.73423424,0.6447993943)
(2.4918633,3.043507787)
(2.5247185,3.613775319)
(6.2241085,6.227302556)
(3.1772487,3.168775554)
(2.103087,2.100544528)
(-6.9860628,-7.713981034)
(-2.522256,-3.049377607)
(1.0516849,-0.1186507342)
(0.62011845,0.8285592264)
(1.5289395,1.928751522)
(2.6504644,2.597750855)
(1.8716765,1.874407033)
(0.13910762,-0.5793155228)
(-3.1521339,-3.488641609)
(-1.4485584,-1.347014267)
(-0.31472012,-0.2100504513)
(4.5880613,4.380362597)
(-0.46920411,-0.6533840644)
(0.92266952,0.928592673)
(1.6010318,1.515413691)
(2.4366902,2.569650651)
(0.71539706,0.7219624529)
(0.13766598,0.2271618016)
(-0.7011098,-0.7670942464)
(3.0553713,3.404113425)
(3.3837306,3.312780234)
(4.5424583,5.652752942)
(2.7351927,2.822547522)
(0.56240961,0.3319228698)
(-5.0532198,-5.550454626)
(-2.1660006,-1.964540901)
(1.1387163,1.266793857)
(3.6765033,3.728177333)
(0.34774383,2.3246242)
(1.5880314,1.903105687)
(-3.2264744,-4.164955146)
(3.0144732,2.829551131)
(-1.3903513,-1.225160316)
(-0.77936547,-1.186137389)
(-0.8858083,-0.860518297)
(3.7754681,3.996892945)
(-1.7442468,-4.356528075)
(1.5045512,1.429786818)
(-4.0766924,-3.647906466)
(1.3905412,1.221821987)
(1.7133596,1.656455391)
(2.0919818,2.251437477)
(0.76173501,0.2519800511)
(1.2561995,1.654029425)
(2.2065374,3.207380407)
(4.3847984,4.241680718)
(1.5160729,1.585759065)
(3.0212964,3.315799803)
(-3.5704474,-2.985417119)
(-1.6857471,-3.564282479)
(0.66962438,-0.03378114412)
(0.75359994,1.085272572)
(2.3587036,2.49549392)
(2.8669731,2.851758483)
(2.3905508,2.279923634)
(0.7345951,0.6475991465)
(-2.8622779,-2.668855848)
(-1.5068216,-1.560226485)
(1.270831,1.174388977)
(3.3901609,3.535906971)
(0.13698987,0.2509355922)
(1.0597115,1.066294965)
(-0.19970384,-0.1027430713)
(2.2570478,2.099845296)
(0.35660409,0.3445741847)
(0.39798124,0.3659899104)
(-1.4234768,-1.271819126)
(5.4346677,4.988241639)
(-0.71921907,-1.195019001)
(0.67131464,0.6723424798)
(-6.6121781,-8.623521008)
(1.721765,2.622155285)
(-0.15869142,-0.1453259876)
(0.3097807,0.3059731711)
(0.67181766,0.6067975927)
(2.5536793,3.117177005)
(2.6420521,3.572010479)
(6.1868836,6.305361623)
(3.1766004,3.1539826)
(2.2267691,2.228898772)
(-6.9429224,-7.48491991)
(-2.4776164,-3.170514927)
(-1.9275462,-2.450275547)
(5.8046927,5.322389826)
(-2.1960387,-3.237002844)
(-0.33812702,-0.3412014784)
(0.39422519,0.8149616227)
(1.0602159,0.8818173728)
(1.4488755,1.483154852)
(2.1435572,3.423971083)
(-0.015815169,-0.8696203184)
(2.4329152,2.528036198)
(1.2551125,3.611744321)
(4.5759095,4.440151307)
(3.1970015,3.132865846)
(1.4175793,1.545226576)
(-4.7446195,-4.397168236)
(-2.2057429,-2.500058294)
(0.29469367,0.2601884336)
(1.3993314,1.960560282)
(3.2496731,3.524574349)
(4.5009998,4.595450024)
(0.49235956,0.8923374744)
(0.621337,0.3452980641)
(-4.8695743,-7.336112702)
(-1.2566835,-0.9950330956)
(-1.7924162,-1.508230216)
(5.4719749,5.444056177)
(0.39176223,0.7306525506)
(1.6763808,1.729688825)
(-0.2637126,-0.8167998405)
(2.6692994,2.72637919)
(0.34221613,0.5395838301)
(-0.78115085,-2.394344905)
(0.021716679,-0.3867616969)
(1.5584533,1.935663323)
(2.8367885,2.822494706)
(3.4559929,3.774621486)
(2.5538554,2.742160312)
(1.2189605,1.130384479)
(-6.3414774,-7.838232297)
(-2.3816767,-2.160217442)
(-0.81506087,-0.5341982222)
(4.4380142,4.808728767)
(0.71485594,2.152724529)
(2.5936173,2.723648342)
(-1.1310039,-1.411958491)
(2.730388,2.564102485)
(-0.19528012,-0.1076444589)
(-0.37183725,-1.060970455)
(-1.235324,-1.201650164)
(4.8564215,4.65962181)
(-1.3809953,-2.734225271)
(0.78231067,0.714979282)
(-2.7199545,-2.119853522)
(0.81672792,0.5454298023)
(2.4620506,2.390676423)
(3.0878808,3.753654574)
(0.69748078,-0.590593932)
(1.3972596,1.49724976)
(1.1618105,2.220871959)
(3.0674561,3.077863593)
(2.8312751,2.83326749)
(2.6350398,2.685287167)
(-5.7816763,-5.40758496)
(-2.1669385,-3.471384014)
(-2.2583178,-2.717537011)
(6.2876794,5.89616831)
(-0.90850198,-1.302302916)
(0.12144266,0.119833805)
(-5.0396556,-5.68840204)
(2.3015025,2.300880059)
(-0.20664784,-0.1679739927)
(0.19684906,0.1320111864)
(1.6135948,1.460699385)
(2.8241681,3.087835346)
(1.8013594,2.804415924)
(6.2482599,6.866401057)
(3.5817251,3.587146749)
(1.2779521,1.283636362)
(-5.9412234,-5.968219472)
(-2.5218195,-2.552930586)
(0.78836959,0.2853601014)
(0.79505188,1.223939641)
(2.5597014,3.005705582)
(3.7479391,3.665532919)
(1.8064815,1.73724136)
(0.21002785,-0.388388298)
(-2.0159415,-2.02727835)
(-1.0086102,-0.9536024185)
(-0.12190236,-0.09356192674)
(3.7827402,4.000496518)
(-0.62603236,-1.931558864)
(1.6056361,1.507961046)
(0.48338887,0.3087209175)
(2.525421,2.670960356)
(0.84117336,0.8661780252)
(-0.30298927,-1.122907585)
(-1.3892537,-2.055157352)
(4.5855088,4.477432334)
(-3.2639999,-5.35993361)
(-0.2233504,-0.1629446001)
(-1.0610388,-0.7024925994)
(1.7579309,1.521677183)
(0.8229545,0.7717110375)
(1.2634209,1.612873822)
(0.53842058,0.2325804977)
(2.1040146,2.502023695)
(2.0083557,3.785834446)
(5.5949791,5.284734995)
(1.9828527,1.918359383)
(1.8258525,2.168762546)
(-3.0909657,-2.684231906)
(-1.6448566,-2.437421204)
(-0.094006949,0.04224529331)
(2.5615203,3.133484902)
(3.4097376,3.627549744)
(5.4311991,5.985249068)
(0.92709691,1.141959485)
(0.18353592,-0.5299747744)
(-3.9069979,-5.227006571)
(-1.1721237,-1.04275014)
(-0.4404266,-0.2291794806)
(4.8251493,4.539219729)
(-0.12020484,0.3281917835)
(1.0525482,1.159564811)
(-1.5986406,-3.008842311)
(3.2328581,3.109755718)
(-1.9014601,-1.512254768)
(-1.3164665,-2.642344609)
(1.0916467,-0.2318170899)
(0.57478862,0.7103726036)
(1.0839987,1.190914301)
(1.5428099,1.502772043)
(3.1147481,3.073438355)
(0.71285062,0.5947124135)
(-4.5349094,-4.47960128)
(-2.1747804,-2.084304536)
(1.2395561,1.260910974)
(3.7870601,3.695139497)
(0.098402643,1.129982314)
(0.8555524,1.004323784)
(1.3445059,1.493660981)
(1.9229019,1.835878238)
(0.54535447,0.5793319356)
(0.84457007,1.526777131)
(-2.2862754,-2.634166474)
(6.2543624,5.920189663)
(-0.634595,-0.9597335366)
(0.34238804,0.339429715)
(-5.1524559,-5.647205541)
(2.155939,2.156494891)
(0.11699825,0.1406061925)
(0.4673683,0.4171053288)
(0.56895273,0.5057641094)
(1.4461325,2.035336088)
(2.9365745,3.655027778)
(5.0425071,4.951162809)
(1.791004,1.645949093)
(2.8028392,3.020468087)
(-6.6393211,-8.788729565)
(-2.0441601,-3.877127759)
(-0.88743665,-1.504461719)
(3.4006958,3.554089342)
(-3.5013317,-6.805730899)
(0.20802865,0.3284452833)
(-2.0648139,-1.963230732)
(2.2645833,1.989398308)
(0.22797265,0.2063649355)
(0.50196055,0.4123572713)
(0.51170699,-0.1675363139)
(1.8249648,2.088290801)
(1.7070116,3.335135634)
(4.5601583,4.401116189)
(2.6371888,2.613883424)
\psset{linewidth=0.3pt,linestyle=solid,linecolor=black}

\psline(-9,-9)(-9,-8.85)    \psline(-9,7)(-9,6.85)  \psline(-9,-9)(-8.85,-9)    \psline(7,-9)(6.85,-9)
\psline(-7,-9)(-7,-8.85)    \psline(-7,7)(-7,6.85)  \psline(-9,-7)(-8.85,-7)    \psline(7,-7)(6.85,-7)
\psline(-5,-9)(-5,-8.85)    \psline(-5,7)(-5,6.85)  \psline(-9,-5)(-8.85,-5)    \psline(7,-5)(6.85,-5)
\psline(-3,-9)(-3,-8.85)    \psline(-3,7)(-3,6.85)  \psline(-9,-3)(-8.85,-3)    \psline(7,-3)(6.85,-3)
\psline(-1,-9)(-1,-8.85)    \psline(-1,7)(-1,6.85)  \psline(-9,-1)(-8.85,-1)    \psline(7,-1)(6.85,-1)
\psline(1,-9)(1,-8.85)  \psline(1,7)(1,6.85)    \psline(-9,1)(-8.85,1)  \psline(7,1)(6.85,1)
\psline(3,-9)(3,-8.85)  \psline(3,7)(3,6.85)    \psline(-9,3)(-8.85,3)  \psline(7,3)(6.85,3)
\psline(5,-9)(5,-8.85)  \psline(5,7)(5,6.85)    \psline(-9,5)(-8.85,5)  \psline(7,5)(6.85,5)
\psline(7,-9)(7,-8.85)  \psline(7,7)(7,6.85)    \psline(-9,7)(-8.85,7)  \psline(7,7)(6.85,7)

\psline(-9,-9)(-9,7)(7,7)(7,-9)(-9,-9)

\psline[linestyle=dashed](-9,-9)(7,7)

\rput(-10.1,-9){$-9$}   \rput(-9,-10.1){$-9$}
\rput(-10.1,-5){$-5$}   \rput(-5,-10.1){$-5$}
\rput(-10.1,-1){$-1$}   \rput(-1,-10.1){$-1$}
\rput(-10.1,3){$3$} \rput(3,-10.1){$3$}
\rput(-10.1,7){$7$} \rput(7,-10.1){$7$}

\rput(0.5,-11.5){$\hat y_{kr}$}
\rput(-11.5,0.5){$\hat y$}

\rput(-0.5,9.5){(b)}

\end{pspicture}
\caption{Smooth supersaturated predictions $\hat y$ against spline
$\hat y_{sp}$ and kriging predictions $\hat y_{kr}$ for the extra
design points in Section \ref{ex2}.
} \label{fig_2d}

\end{center}
\end{figure}
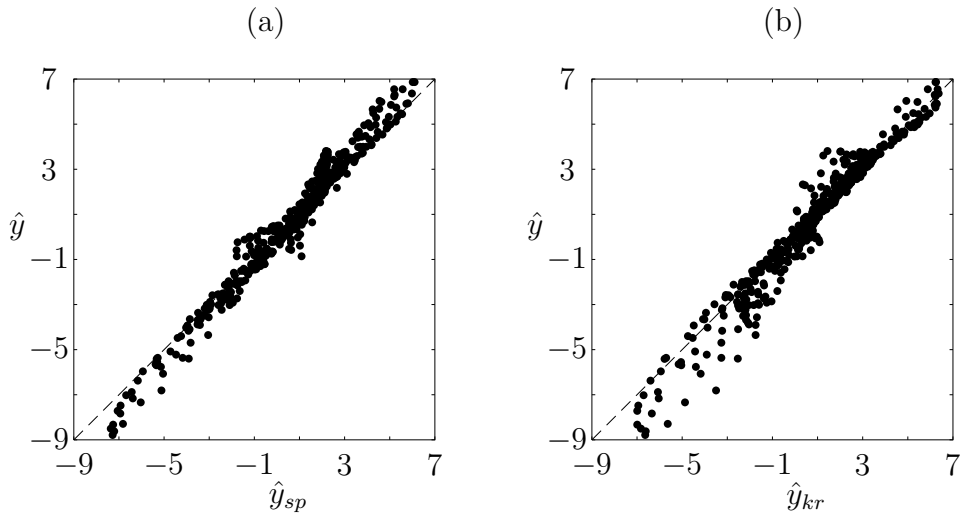

\begin{figure}[h!]  
\begin{center}
\psset{unit=3.0mm,linewidth=0.85pt}
 \begin{pspicture}(-10,-12)(10,11)
     \put(0,0){}

\psset{linecolor=black}

\psdots[dotsize=3pt] (0.98636205,2.627844873)
(2.509042909,3.775017232) (7.023502804,5.678613789)
(2.496513133,2.398787689) (2.386586264,2.66753671)
(-5.740675877,-8.325055083) (-1.764218965,-3.50859752)
(-0.168430224,-1.282328189) (4.132001604,3.98052527)
(-1.468491645,-5.397169546) (-0.598098158,0.89090631)
(-3.169107714,-2.590265361) (1.918636123,1.49826681)
(0.999535919,0.9605137703) (1.114740974,1.554615297)
(-0.197564846,-0.4868283817) (2.401995863,2.575943271)
(1.03903931,3.805403597) (6.94197029,4.871757466)
(2.03024014,2.742507468) (1.054519578,1.528787946)
(-4.021225072,-3.659410646) (-1.477331183,-2.31140841)
(0.092594171,-0.5578475178) (2.612309542,3.519012498)
(3.575348652,3.411621777) (3.953401796,6.034284451)
(3.068540578,2.273504707) (0.556268895,-0.09690062294)
(-3.982184083,-5.076233684) (-2.686513892,-1.670028519)
(-1.927609008,-1.192354622) (5.808369411,5.116926356)
(0.013804769,-0.01416654384) (0.955361808,1.183009218)
(-0.735410796,-1.588246607) (3.078524452,2.807541421)
(0.594201807,-0.1765101336) (-0.287440413,-2.855657844)
(-1.122338401,-3.023353796) (8.019167629,5.73463237)
(-1.838962635,-2.780515147) (-1.366758466,-0.6573419382)
(-3.428010951,-5.760788934) (2.94024641,2.894762614)
(-1.276526235,-1.448205123) (-0.446418637,-0.9733922831)
(1.157187722,0.8453479166) (2.065472516,3.193242054)
(1.957173202,3.33854127) (6.947895432,6.543790194)
(3.406953915,3.488591733) (1.766533864,1.82223938)
(-6.401371923,-7.02422391) (-2.374883527,-2.881888781)
(0.289934545,0.4795575467) (0.253073491,1.822306389)
(2.530687687,3.546882613) (5.903635804,4.66178512)
(1.178745048,1.10723608) (0.158271769,-0.8616599753)
(-0.610312931,-0.6380465386) (-0.091980244,-0.04566742967)
(0.465933931,0.3393841142) (3.936377998,4.00240948)
(-0.244276121,-0.5496043174) (1.144622722,1.123148026)
(0.746007152,0.7448846056) (2.375258447,2.414442117)
(0.662890147,0.5653902311) (0.20301397,-0.05447153561)
(0.139660524,0.03217984645) (0.243301625,1.538079273)
(2.735884388,3.039312256) (3.734350536,3.701772003)
(1.85562568,1.448785053) (1.196478858,0.8684140376)
(-3.876484646,-8.292449657) (-2.903623197,-1.487598819)
(-1.29726748,0.5917251247) (3.406995366,4.107423576)
(0.292707579,2.289612134) (1.732847233,2.282535172)
(-2.348218294,-3.044032638) (3.211222771,2.654719085)
(-0.241034646,-0.7385731162) (-0.078886725,-1.044446335)
(-0.921139955,-1.961234691) (6.984706961,5.088436094)
(-1.256230372,-3.019964179) (-0.624375784,0.08848158648)
(-0.259081599,-0.2572211669) (1.090164001,0.6817004083)
(1.746482933,1.869121677) (1.404141701,3.662506523)
(0.094841936,-0.06039267212) (0.421530038,1.207182974)
(1.422791844,2.458275517) (4.321491363,3.303710171)
(0.157816738,1.788463848) (2.967198657,3.393995204)
(-4.724050459,-3.92975832) (-1.008550627,-3.903484432)
(-0.57786028,-2.548195214) (7.043218124,4.972606857)
(-2.475010141,-4.691973254) (-1.938584127,-0.6534973466)
(0.382636754,0.5090169776) (1.74300292,1.403772238)
(0.589461673,0.8677802806) (0.746312126,2.129790571)
(-0.1661808,-0.8653304294) (1.833225075,2.042280526)
(0.88363748,2.826105562) (4.468444574,3.594842016)
(2.419371571,3.28268103) (1.805128462,2.110867934)
(-5.958874978,-5.363115189) (-1.830865837,-3.0098337)
(0.345894513,0.2542782036) (0.849117234,2.629523333)
(3.180286212,3.721020467) (5.598611727,5.485603547)
(1.637564775,0.7835269699) (0.540844407,-0.255734196)
(-3.066289367,-6.063753316) (-2.477777397,-0.8375556905)
(-2.726958328,-1.229149434) (4.012631686,5.322537075)
(0.711781033,1.828409882) (2.976311458,2.645406587)
(-0.546220809,-0.6542220126) (2.87567613,2.652432806)
(1.001052531,0.3414910559) (0.013209181,-1.440854112)
(0.127480725,0.01509667654) (-0.020469401,0.8693858149)
(1.837003615,2.313617471) (3.277153158,2.698522859)
(2.604853802,2.405133296) (0.48168212,0.1147154057)
(-3.211700058,-3.161531568) (-1.730019501,-1.5755806)
(1.824675028,1.572272665) (2.483373404,3.093036599)
(-0.067224663,0.4480248846) (1.007572388,1.00087844)
(-1.980248294,-1.704649265) (2.555394694,2.026225758)
(0.392538173,0.1835133029) (0.464926847,0.3323405695)
(-1.692126034,-2.055281271) (7.061558866,5.479173874)
(-0.54285416,-1.291587615) (0.050358572,0.4238666351)
(-5.035808685,-7.166426858) (1.890295761,2.380456444)
(0.009245474,-0.0209519814) (0.449520913,0.2983573339)
(0.63806668,0.2447598666) (2.072065701,3.444242221)
(2.888544722,3.420299184) (5.81203582,6.547412916)
(2.026481387,2.684669014) (2.678743101,2.819239168)
(-6.403146004,-6.380083525) (-1.560897268,-3.678124917)
(0.037651842,-0.6356074522) (1.657796084,2.607148473)
(3.111468315,3.051956799) (3.058884586,4.551215978)
(3.672622572,3.120621839) (1.123103706,0.9189461476)
(-5.378737925,-6.879249271) (-2.959998085,-2.216864258)
(-0.481326817,0.679740554) (4.260300717,4.068236195)
(0.027773293,1.173365957) (0.804052368,1.276772403)
(-1.867122514,-4.090677019) (3.568751699,3.220111655)
(-1.766812592,-2.00541582) (-0.680408822,-2.156570941)
(-0.138551395,-0.797884014) (4.559803651,4.326802657)
(-0.717650767,-2.91664476) (0.695601241,1.164840228)
(-4.127539272,-3.363747721) (1.01380554,0.704835473)
(2.425683987,2.455127189) (2.459114663,3.396501974)
(0.27720255,0.351930509) (1.130918016,2.22466658)
(1.85330371,3.651434846) (7.073458682,5.038097601)
(0.050518953,1.62758707) (1.621914451,2.802560631)
(-2.675802157,-2.5720765) (-0.861085865,-2.937331019)
(-1.658171163,-3.216416284) (6.856982115,6.355235231)
(-0.48529501,-0.6910182443) (0.013559086,0.2162232382)
(-3.847056381,-4.020300148) (2.451102067,2.067419937)
(0.329231898,0.1012611267) (0.560207864,0.3609517788)
(0.620206016,0.5286039602) (1.905714462,2.806002971)
(1.578982723,3.790338765) (8.010538802,5.860377931)
(3.092290334,2.78486689) (1.728393941,1.679321266)
(-5.386013631,-8.50803814) (-2.630756624,-2.558203806)
(0.126166106,0.005634908937) (-0.032092887,0.8383188922)
(1.733847948,2.244289325) (3.292990795,2.67056069)
(2.638533837,2.358692684) (0.45578341,-0.02159357759)
(-3.227564932,-3.250950615) (-1.827668891,-1.556634673)
(-1.133039297,-1.093987302) (6.126340018,4.870010287)
(-0.425795869,-1.214166877) (0.437111502,0.7066807478)
(1.877128052,1.617425486) (2.390423569,2.984940031)
(0.952835186,0.959494669) (0.086330545,0.05569085578)
(-1.40169088,-3.803050039) (5.923524268,6.866469785)
(-0.220471792,-0.2219034312) (0.28556768,0.2934501506)
(-2.402988133,-2.1319896) (2.520477658,1.987697577)
(0.410437931,0.2000372253) (0.521061918,0.4154171654)
(0.858210966,0.6447993943) (1.685668844,3.043507787)
(2.148481313,3.613775319) (7.337205117,6.227302556)
(3.178926874,3.168775554) (2.013292113,2.100544528)
(-6.276922271,-7.713981034) (-2.240737704,-3.049377607)
(0.100169823,-0.1186507342) (0.081245395,0.8285592264)
(1.311254167,1.928751522) (3.263035819,2.597750855)
(2.427632425,1.874407033) (0.348874686,-0.5793155228)
(-2.962656922,-3.488641609) (-2.022222115,-1.347014267)
(-0.344468213,-0.2100504513) (4.981237283,4.380362597)
(-0.286595974,-0.6533840644) (0.814704267,0.928592673)
(1.699346543,1.515413691) (2.237238781,2.569650651)
(0.661330336,0.7219624529) (0.161300522,0.2271618016)
(0.011552507,-0.7670942464) (2.677465641,3.404113425)
(3.531741837,3.312780234) (3.41627771,5.652752942)
(3.489893177,2.822547522) (0.737935757,0.3319228698)
(-4.673243957,-5.550454626) (-2.818973957,-1.964540901)
(-0.443257368,1.266793857) (3.324870522,3.728177333)
(0.170747998,2.3246242) (1.07040523,1.903105687)
(-2.750196532,-4.164955146) (3.255716336,2.829551131)
(-0.846355669,-1.225160316) (-0.291437633,-1.186137389)
(-0.047204077,-0.860518297) (3.745735697,3.996892945)
(-0.958636608,-4.356528075) (0.433917625,1.429786818)
(-4.342610684,-3.647906466) (1.490311519,1.221821987)
(1.648874893,1.656455391) (1.756443135,2.251437477)
(0.178601765,0.2519800511) (0.601859698,1.654029425)
(1.825127149,3.207380407) (5.805136645,4.241680718)
(-0.250102182,1.585759065) (2.392550754,3.315799803)
(-3.472495789,-2.985417119) (-0.894788693,-3.564282479)
(0.115620494,-0.03378114412) (0.145439324,1.085272572)
(2.190580893,2.49549392) (3.107831041,2.851758483)
(2.159696517,2.279923634) (0.568607204,0.6475991465)
(-2.776481395,-2.668855848) (-1.234732487,-1.560226485)
(1.293142007,1.174388977) (3.119538056,3.535906971)
(-0.121736993,0.2509355922) (1.091293248,1.066294965)
(-0.31106951,-0.1027430713) (2.431215141,2.099845296)
(0.417736521,0.3445741847) (0.360590039,0.3659899104)
(-1.313273527,-1.271819126) (6.337341491,4.988241639)
(-0.42467819,-1.195019001) (0.37674171,0.6723424798)
(-4.893078315,-8.623521008) (1.437493391,2.622155285)
(-0.160545927,-0.1453259876) (0.338191835,0.3059731711)
(0.855523576,0.6067975927) (1.698624598,3.117177005)
(2.298556983,3.572010479) (7.076037444,6.305361623)
(3.043816571,3.1539826) (2.118395488,2.228898772)
(-6.403405222,-7.48491991) (-2.113537391,-3.170514927)
(-0.979707078,-2.450275547) (7.58462706,5.322389826)
(-1.664943071,-3.237002844) (-1.175468666,-0.3412014784)
(1.023500483,0.8149616227) (1.236320145,0.8818173728)
(1.177351379,1.483154852) (1.002376528,3.423971083)
(-0.324770473,-0.8696203184) (2.484404095,2.528036198)
(0.901078736,3.611744321) (6.003805131,4.440151307)
(2.480543465,3.132865846) (1.199418491,1.545226576)
(-4.889664196,-4.397168236) (-1.785116505,-2.500058294)
(0.23008762,0.2601884336) (0.299173535,1.960560282)
(3.025601961,3.524574349) (4.885654324,4.595450024)
(1.488318774,0.8923374744) (0.833006175,0.3452980641)
(-3.373254829,-7.336112702) (-2.746528298,-0.9950330956)
(-2.452545728,-1.508230216) (5.28413659,5.444056177)
(0.381230568,0.7306525506) (1.722304662,1.729688825)
(-0.474868289,-0.8167998405) (2.834785675,2.72637919)
(1.419589376,0.5395838301) (-0.124011672,-2.394344905)
(0.070885806,-0.3867616969) (0.880409878,1.935663323)
(2.757022507,2.822494706) (2.988197274,3.774621486)
(3.251855066,2.742160312) (1.322495148,1.130384479)
(-5.112299656,-7.838232297) (-2.962672124,-2.160217442)
(-2.36166389,-0.5341982222) (3.579680172,4.808728767)
(0.558749175,2.152724529) (2.813040182,2.723648342)
(-1.265627518,-1.411958491) (3.037483221,2.564102485)
(0.459793492,-0.1076444589) (0.047841816,-1.060970455)
(-0.530084695,-1.201650164) (5.668954483,4.65962181)
(-0.836381328,-2.734225271) (0.222959247,0.714979282)
(-2.600895683,-2.119853522) (0.956652274,0.5454298023)
(2.404694354,2.390676423) (2.150708995,3.753654574)
(-0.031597642,-0.590593932) (1.081169014,1.49724976)
(0.9435566,2.220871959) (3.766241868,3.077863593)
(1.737439562,2.83326749) (2.409210064,2.685287167)
(-6.058231438,-5.40758496) (-1.530672792,-3.471384014)
(-1.671663019,-2.717537011) (7.376713221,5.89616831)
(-0.736243411,-1.302302916) (-0.265740929,0.119833805)
(-4.522731323,-5.68840204) (2.332433352,2.300880059)
(-0.008596806,-0.1679739927) (0.376706395,0.1320111864)
(1.740498122,1.460699385) (2.408920373,3.087835346)
(1.494326218,2.804415924) (6.060759272,6.866401057)
(3.565372908,3.587146749) (1.269998369,1.283636362)
(-5.915848772,-5.968219472) (-2.498674536,-2.552930586)
(0.186269534,0.2853601014) (-0.035836854,1.223939641)
(2.254536733,3.005705582) (4.589637352,3.665532919)
(1.968472437,1.73724136) (0.291976512,-0.388388298)
(-1.979523865,-2.02727835) (-1.121937585,-0.9536024185)
(0.370605162,-0.09356192674) (3.55446153,4.000496518)
(-0.399640295,-1.931558864) (1.396683568,1.507961046)
(0.348903987,0.3087209175) (2.471160099,2.670960356)
(1.337002629,0.8661780252) (0.074640673,-1.122907585)
(-0.376890608,-2.055157352) (5.780440593,4.477432334)
(-2.218223931,-5.35993361) (-1.695183759,-0.1629446001)
(-1.116762296,-0.7024925994) (1.953068261,1.521677183)
(0.673893403,0.7717110375) (0.817273109,1.612873822)
(0.25074521,0.2325804977) (1.692737308,2.502023695)
(1.56220409,3.785834446) (7.58624883,5.284734995)
(0.763552209,1.918359383) (1.188628957,2.168762546)
(-2.736974038,-2.684231906) (-0.945348152,-2.437421204)
(0.358679851,0.04224529331) (1.557616273,3.133484902)
(3.309534123,3.627549744) (5.322276902,5.985249068)
(2.073240146,1.141959485) (0.437026259,-0.5299747744)
(-3.094954445,-5.227006571) (-2.416620743,-1.04275014)
(-1.146902168,-0.2291794806) (5.102757293,4.539219729)
(-0.037229531,0.3281917835) (0.858928508,1.159564811)
(-1.279737017,-3.008842311) (3.507900492,3.109755718)
(-1.117976968,-1.512254768) (-0.546598359,-2.642344609)
(0.080914643,-0.2318170899) (0.290122544,0.7103726036)
(0.975225966,1.190914301) (1.7731432,1.502772043)
(3.219893901,3.073438355) (0.755684854,0.5947124135)
(-4.572419177,-4.47960128) (-2.268532317,-2.084304536)
(0.637171698,1.260910974) (3.772192722,3.695139497)
(-0.054472268,1.129982314) (0.669646956,1.004323784)
(1.572692755,1.493660981) (1.969489004,1.835878238)
(0.229635577,0.5793319356) (0.410957258,1.526777131)
(-1.809896269,-2.634166474) (7.129071243,5.920189663)
(-0.497381311,-0.9597335366) (0.02925847,0.339429715)
(-4.751851583,-5.647205541) (2.179467314,2.156494891)
(0.273655425,0.1406061925) (0.605286841,0.4171053288)
(0.336409526,0.5057641094) (0.408382319,2.035336088)
(2.563965042,3.655027778) (6.263818793,4.951162809)
(1.894944148,1.645949093) (2.439626698,3.020468087)
(-4.828281798,-8.788729565) (-1.359386763,-3.877127759)
(-0.133183214,-1.504461719) (3.722010881,3.554089342)
(-2.062586337,-6.805730899) (-1.593566798,0.3284452833)
(-2.247443969,-1.963230732) (2.521617769,1.989398308)
(0.410173496,0.2063649355) (0.510253992,0.4123572713)
(0.01429945,-0.1675363139) (1.531296265,2.088290801)
(1.329605092,3.335135634) (6.134253773,4.401116189)
(1.289002319,2.613883424)

\psset{linewidth=0.3pt,linestyle=solid,linecolor=black}

\psline(-9,-9)(-9,-8.85)    \psline(-9,7)(-9,6.85)
\psline(-9,-9)(-8.85,-9)    \psline(8.5,-9)(8.35,-9)
\psline(-7,-9)(-7,-8.85)    \psline(-7,7)(-7,6.85)
\psline(-9,-7)(-8.85,-7)    \psline(8.5,-7)(8.35,-7)
\psline(-5,-9)(-5,-8.85)    \psline(-5,7)(-5,6.85)
\psline(-9,-5)(-8.85,-5)    \psline(8.5,-5)(8.35,-5)
\psline(-3,-9)(-3,-8.85)    \psline(-3,7)(-3,6.85)
\psline(-9,-3)(-8.85,-3)    \psline(8.5,-3)(8.35,-3)
\psline(-1,-9)(-1,-8.85)    \psline(-1,7)(-1,6.85)
\psline(-9,-1)(-8.85,-1)    \psline(8.5,-1)(8.35,-1)
\psline(1,-9)(1,-8.85)  \psline(1,7)(1,6.85) \psline(-9,1)(-8.85,1)
\psline(8.5,1)(8.35,1) \psline(3,-9)(3,-8.85) \psline(3,7)(3,6.85)
\psline(-9,3)(-8.85,3)  \psline(8.5,3)(8.35,3)
\psline(5,-9)(5,-8.85) \psline(5,7)(5,6.85) \psline(-9,5)(-8.85,5)
\psline(8.5,5)(8.35,5) \psline(7,-9)(7,-8.85) \psline(7,7)(7,6.85)
\psline(-9,7)(-8.85,7) \psline(8.5,7)(8.35,7)

\psline(-9,-9)(-9,7)(8.5,7)(8.5,-9)(-9,-9)

\psline[linestyle=dashed](-9,-9)(7,7)

\rput(-10.1,-9){$-9$}   \rput(-9,-10.1){$-9$}
\rput(-10.1,-5){$-5$}   \rput(-5,-10.1){$-5$}
\rput(-10.1,-1){$-1$}   \rput(-1,-10.1){$-1$}
\rput(-10.1,3){$3$} \rput(3,-10.1){$3$}
\rput(-10.1,7){$7$} \rput(7,-10.1){$7$}

\rput(0.5,-11.5){$ y$} \rput(-11.5,0.5){$\hat y$}
\rput(-0.5,9.5){(a)}

\end{pspicture}
 \begin{pspicture}(-12.5,-12)(7.5,11)
     \put(0,0){}

\psset{linecolor=black}

\psdots[dotsize=3pt] (0.98636205,2.52323973)
(2.509042909,2.24275589) (7.023502804,4.54063436)
(2.496513133,2.15297424) (2.386586264,2.64242987)
(-5.740675877,-7.26006482) (-1.764218965,-3.38843814)
(-0.168430224,-1.11940195) (4.132001604,3.59206185)
(-1.468491645,-3.90553369) (-0.598098158,0.95921724)
(-3.169107714,-2.85985271) (1.918636123,1.64605616)
(0.999535919,1.10099795) (1.114740974,1.45646942)
(-0.197564846,-0.82960019) (2.401995863,2.17538763)
(1.03903931,2.1689318) (6.94197029,4.14700148)
(2.03024014,2.52587097) (1.054519578,1.9207745)
(-4.021225072,-3.39131079) (-1.477331183,-2.14136371)
(0.092594171,-0.48464051) (2.612309542,3.26995646)
(3.575348652,2.23627324) (3.953401796,4.55431819)
(3.068540578,2.12158154) (0.556268895,0.61617251)
(-3.982184083,-4.72028519) (-2.686513892,-1.27230106)
(-1.927609008,-1.10138865) (5.808369411,5.26245867)
(0.013804769,0.27953291) (0.955361808,1.3099938)
(-0.735410796,-1.9014544) (3.078524452,2.1642379)
(0.594201807,-0.04812015) (-0.287440413,-2.06885211)
(-1.122338401,-2.58611967) (8.019167629,5.29269757)
(-1.838962635,-2.28705212) (-1.366758466,-0.17841159)
(-3.428010951,-5.14108371) (2.94024641,2.11448315)
(-1.276526235,-0.98709301) (-0.446418637,-0.72193718)
(1.157187722,1.12472685) (2.065472516,2.97271084)
(1.957173202,2.16329264) (6.947895432,5.57627079)
(3.406953915,3.17386288) (1.766533864,1.77999476)
(-6.401371923,-6.68535107) (-2.374883527,-2.96345818)
(0.289934545,-0.14920712) (0.253073491,1.98470821)
(2.530687687,2.09459688) (5.903635804,3.63238471)
(1.178745048,1.06029187) (0.158271769,1.09463844)
(-0.610312931,-0.57717082) (-0.091980244,0.01431977)
(0.465933931,0.40581339) (3.936377998,3.96722772)
(-0.244276121,-0.43469921) (1.144622722,1.09226509)
(0.746007152,0.78426102) (2.375258447,2.36685182)
(0.662890147,0.55538898) (0.20301397,0.0370783)
(0.139660524,-0.88270572) (0.243301625,1.78248217)
(2.735884388,1.98582349) (3.734350536,2.98265986)
(1.85562568,1.00363086) (1.196478858,1.18098198)
(-3.876484646,-6.82192885) (-2.903623197,-1.38664247)
(-1.29726748,0.53743671) (3.406995366,4.19581582)
(0.292707579,1.45522352) (1.732847233,2.19687137)
(-2.348218294,-2.99459391) (3.211222771,2.14425283)
(-0.241034646,-0.23889504) (-0.078886725,-0.67233904)
(-0.921139955,-1.58142314) (6.984706961,4.82835045)
(-1.256230372,-2.09031973) (-0.624375784,0.4670119)
(-0.259081599,-0.59116056) (1.090164001,1.12418248)
(1.746482933,1.60576734) (1.404141701,2.75586901)
(0.094841936,-1.20333293) (0.421530038,1.40057739)
(1.422791844,1.56755193) (4.321491363,2.65286827)
(0.157816738,1.56242477) (2.967198657,3.45167429)
(-4.724050459,-3.96851593) (-1.008550627,-3.40630217)
(-0.57786028,-2.11814128) (7.043218124,4.37215696)
(-2.475010141,-3.82240929) (-1.938584127,-0.20215006)
(0.382636754,0.11246408) (1.74300292,1.6416825)
(0.589461673,0.98806439) (0.746312126,1.63648262)
(-0.1661808,-1.78098887) (1.833225075,1.77136432)
(0.88363748,1.78321232) (4.468444574,3.01896587)
(2.419371571,3.04994038) (1.805128462,2.08984374)
(-5.958874978,-5.27896951) (-1.830865837,-2.9692101)
(0.345894513,0.05970666) (0.849117234,2.5952268)
(3.180286212,2.24625639) (5.598611727,4.26163964)
(1.637564775,0.68370111) (0.540844407,0.62774397)
(-3.066289367,-5.04757319) (-2.477777397,-0.49143293)
(-2.726958328,-1.14388645) (4.012631686,5.34230311)
(0.711781033,1.32221823) (2.976311458,2.30428534)
(-0.546220809,-0.70402922) (2.87567613,2.30333804)
(1.001052531,0.51390755) (0.013209181,-0.92261659)
(0.127480725,-1.18036381) (-0.020469401,1.22539813)
(1.837003615,1.47536474) (3.277153158,2.17518847)
(2.604853802,2.21182759) (0.48168212,1.03777478)
(-3.211700058,-2.98060636) (-1.730019501,-1.3078123)
(1.824675028,1.6044876) (2.483373404,2.98543681)
(-0.067224663,-0.21972459) (1.007572388,1.00104176)
(-1.980248294,-1.87334819) (2.555394694,1.97733455)
(0.392538173,0.52213082) (0.464926847,0.45219493)
(-1.692126034,-1.85767786) (7.061558866,5.48108008)
(-0.54285416,-0.72850048) (0.050358572,0.73323145)
(-5.035808685,-6.39050659) (1.890295761,1.89810675)
(0.009245474,0.11293003) (0.449520913,0.38166725)
(0.63806668,0.53725579) (2.072065701,3.22400994)
(2.888544722,2.2175756) (5.81203582,5.20141769)
(2.026481387,2.34367278) (2.678743101,2.81386607)
(-6.403146004,-6.16181009) (-1.560897268,-3.50408949)
(0.037651842,-1.11668036) (1.657796084,2.5106742)
(3.111468315,2.072021) (3.058884586,3.52770209)
(3.672622572,2.83359987) (1.123103706,1.05124192)
(-5.378737925,-6.43638768) (-2.959998085,-2.14243387)
(-0.481326817,0.58006022) (4.260300717,4.18198604)
(0.027773293,0.80257096) (0.804052368,1.43080643)
(-1.867122514,-3.87445557) (3.568751699,2.22317664)
(-1.766812592,-1.55565357) (-0.680408822,-1.75178112)
(-0.138551395,-0.46893017) (4.559803651,4.09725399)
(-0.717650767,-1.95431284) (0.695601241,1.17926363)
(-4.127539272,-3.49104934) (1.01380554,1.20592497)
(2.425683987,1.84374908) (2.459114663,2.78733252)
(0.27720255,-0.04424975) (1.130918016,2.13602064)
(1.85330371,2.14567018) (7.073458682,4.05626988)
(0.050518953,1.51891107) (1.621914451,3.09778647)
(-2.675802157,-2.30050073) (-0.861085865,-2.47152154)
(-1.658171163,-3.00323512) (6.856982115,5.98194215)
(-0.48529501,-0.60151211) (0.013559086,0.36474687)
(-3.847056381,-4.02865163) (2.451102067,1.8796919)
(0.329231898,0.49508949) (0.560207864,0.54152808)
(0.620206016,0.63916858) (1.905714462,2.54918798)
(1.578982723,2.23986497) (8.010538802,5.05132827)
(3.092290334,2.34534128) (1.728393941,1.76346353)
(-5.386013631,-7.36784937) (-2.630756624,-2.62847907)
(0.126166106,-1.20868745) (-0.032092887,1.1919357)
(1.733847948,1.41743919) (3.292990795,2.14479528)
(2.638533837,2.17482662) (0.45578341,0.93314811)
(-3.227564932,-3.0792802) (-1.827668891,-1.25938216)
(-1.133039297,-0.88888509) (6.126340018,4.907575)
(-0.425795869,-0.55955175) (0.437111502,0.93855237)
(1.877128052,1.65006668) (2.390423569,2.87934896)
(0.952835186,0.9603955) (0.086330545,0.13036653)
(-1.40169088,-3.46930047) (5.923524268,6.09977347)
(-0.220471792,-0.2422266) (0.28556768,0.28697544)
(-2.402988133,-2.30029013) (2.520477658,1.93484955)
(0.410437931,0.55606653) (0.521061918,0.5359885)
(0.858210966,0.90178726) (1.685668844,2.81673682)
(2.148481313,2.22815853) (7.337205117,5.19910118)
(3.178926874,2.77944469) (2.013292113,2.09061641)
(-6.276922271,-7.06087896) (-2.240737704,-3.10950464)
(0.100169823,-1.43174268) (0.081245395,1.12021582)
(1.311254167,1.24295968) (3.263035819,2.11633961)
(2.427632425,1.73861691) (0.348874686,0.63159648)
(-2.962656922,-3.2168286) (-2.022222115,-0.90939335)
(-0.344468213,-0.10977551) (4.981237283,4.41331116)
(-0.286595974,-0.29794396) (0.814704267,1.03426869)
(1.699346543,1.52415752) (2.237238781,2.52035023)
(0.661330336,0.72824142) (0.161300522,0.22837756)
(0.011552507,-0.81547066) (2.677465641,3.14436077)
(3.531741837,2.20064607) (3.41627771,4.24743477)
(3.489893177,2.66632561) (0.737935757,0.73695567)
(-4.673243957,-5.30518711) (-2.818973957,-1.6923456)
(-0.443257368,1.10387502) (3.324870522,3.80736306)
(0.170747998,1.38219955) (1.07040523,1.9627278)
(-2.750196532,-3.9203569) (3.255716336,2.14602757)
(-0.846355669,-0.69610013) (-0.291437633,-0.8514659)
(-0.047204077,-0.61365041) (3.745735697,3.6881039)
(-0.958636608,-3.04235382) (0.433917625,1.38556314)
(-4.342610684,-3.86370161) (1.490311519,1.46217441)
(1.648874893,1.50413751) (1.756443135,2.02801091)
(0.178601765,-0.56120912) (0.601859698,1.7637157)
(1.825127149,1.93184414) (5.805136645,3.34765388)
(-0.250102182,1.43229512) (2.392550754,3.41823295)
(-3.472495789,-2.91496033) (-0.894788693,-3.02685693)
(0.115620494,-1.17174013) (0.145439324,1.40109215)
(2.190580893,1.71261047) (3.107831041,2.37573998)
(2.159696517,2.0634495) (0.568607204,1.563824)
(-2.776481395,-2.35369378) (-1.234732487,-1.34414824)
(1.293142007,1.16914684) (3.119538056,3.47336166)
(-0.121736993,-0.12703585) (1.091293248,1.07412577)
(-0.31106951,-0.22838231) (2.431215141,2.11556414)
(0.417736521,0.51083899) (0.360590039,0.39210938)
(-1.313273527,-1.06780888) (6.337341491,5.03681368)
(-0.42467819,-0.54139645) (0.37674171,0.9247984)
(-4.893078315,-7.20402909) (1.437493391,1.95181125)
(-0.160545927,-0.15853056) (0.338191835,0.29216823)
(0.855523576,0.8837675) (1.698624598,2.89380054)
(2.298556983,2.22523401) (7.076037444,5.21501787)
(3.043816571,2.77519949) (2.118395488,2.20831174)
(-6.403405222,-6.93016555) (-2.113537391,-3.20555301)
(-0.979707078,-2.03268228) (7.58462706,4.95629621)
(-1.664943071,-2.44807464) (-1.175468666,0.12779768)
(1.023500483,0.38562949) (1.236320145,1.22399269)
(1.177351379,1.42674895) (1.002376528,2.49862405)
(-0.324770473,-1.36554322) (2.484404095,2.07340117)
(0.901078736,2.08519481) (6.003805131,3.76509892)
(2.480543465,2.95219141) (1.199418491,1.73960311)
(-4.889664196,-4.2514383) (-1.785116505,-2.40381216)
(0.23008762,-0.37589808) (0.299173535,2.11909164)
(3.025601961,2.14301287) (4.885654324,3.58560225)
(1.488318774,0.60015346) (0.833006175,0.87100564)
(-3.373254829,-6.03375927) (-2.746528298,-0.78187555)
(-2.452545728,-1.46598292) (5.28413659,5.54856887)
(0.381230568,0.7080103) (1.722304662,1.66201802)
(-0.474868289,-1.01329708) (2.834785675,2.24936243)
(1.419589376,0.62699372) (-0.124011672,-1.64610264)
(0.070885806,-1.19499376) (0.880409878,2.02422143)
(2.757022507,1.97362829) (2.988197274,3.04753669)
(3.251855066,2.30466875) (1.322495148,1.28373062)
(-5.112299656,-6.94225182) (-2.962672124,-2.13885732)
(-2.36166389,-0.4465487) (3.579680172,4.8665685)
(0.558749175,1.47418406) (2.813040182,2.42469304)
(-1.265627518,-1.47274458) (3.037483221,2.21623404)
(0.459793492,0.22423782) (0.047841816,-0.64125534)
(-0.530084695,-0.86082722) (5.668954483,4.46296239)
(-0.836381328,-1.76444426) (0.222959247,0.88724925)
(-2.600895683,-2.28628454) (0.956652274,1.10172451)
(2.404694354,1.82145017) (2.150708995,2.95934036)
(-0.031597642,-1.78813005) (1.081169014,1.47776147)
(0.9435566,1.50480621) (3.766241868,2.58018842)
(1.737439562,2.52087261) (2.409210064,2.67713572)
(-6.058231438,-5.32678017) (-1.530672792,-3.30223072)
(-1.671663019,-2.50077355) (7.376713221,5.73350813)
(-0.736243411,-0.9443643) (-0.265740929,0.42889737)
(-4.522731323,-5.33465555) (2.332433352,1.91102143)
(-0.008596806,0.17473486) (0.376706395,0.31780025)
(1.740498122,1.54507845) (2.408920373,2.96070546)
(1.494326218,2.00588705) (6.060759272,6.02498485)
(3.565372908,3.55370162) (1.269998369,1.25459667)
(-5.915848772,-5.94159018) (-2.498674536,-2.57675229)
(0.186269534,-0.70725409) (-0.035836854,1.54817538)
(2.254536733,1.82559135) (4.589637352,2.87357205)
(1.968472437,1.5725483) (0.291976512,1.01689214)
(-1.979523865,-1.84515452) (-1.121937585,-0.71114642)
(0.370605162,0.18527468) (3.55446153,3.82297604)
(-0.399640295,-1.40049294) (1.396683568,1.39217948)
(0.348903987,0.38111163) (2.471160099,2.44278247)
(1.337002629,0.85619119) (0.074640673,-0.70509538)
(-0.376890608,-1.75980957) (5.780440593,3.92951237)
(-2.218223931,-4.18107982) (-1.695183759,0.15034967)
(-1.116762296,-1.03281783) (1.953068261,1.71800401)
(0.673893403,0.92600491) (0.817273109,1.36402787)
(0.25074521,0.01353983) (1.692737308,2.27162216)
(1.56220409,2.20086228) (7.58624883,4.39155229)
(0.763552209,1.77298617) (1.188628957,2.6565748)
(-2.736974038,-2.30255361) (-0.945348152,-2.0941405)
(0.358679851,0.09726103) (1.557616273,2.97599158)
(3.309534123,2.25946838) (5.322276902,4.63651496)
(2.073240146,1.08301734) (0.437026259,0.53594382)
(-3.094954445,-4.47306995) (-2.416620743,-0.59928659)
(-1.146902168,-0.19665211) (5.102757293,4.67575928)
(-0.037229531,0.43450719) (0.858928508,1.30761291)
(-1.279737017,-3.10305281) (3.507900492,2.19891239)
(-1.117976968,-1.18558252) (-0.546598359,-2.05992878)
(0.080914643,-1.7432366) (0.290122544,0.94919447)
(0.975225966,0.79234073) (1.7731432,1.26675644)
(3.219893901,2.95962939) (0.755684854,1.00738418)
(-4.572419177,-4.40107862) (-2.268532317,-1.89552551)
(0.637171698,1.08650422) (3.772192722,3.75585834)
(-0.054472268,0.58551404) (0.669646956,1.17849873)
(1.572692755,1.19680744) (1.969489004,1.971242)
(0.229635577,0.73386341) (0.410957258,1.13166878)
(-1.809896269,-2.47261434) (7.129071243,5.81027241)
(-0.497381311,-0.65889594) (0.02925847,0.5922843)
(-4.751851583,-5.37375051) (2.179467314,1.85308114)
(0.273655425,0.43216793) (0.605286841,0.57729743)
(0.336409526,0.03201539) (0.408382319,2.12843385)
(2.563965042,2.15164214) (6.263818793,3.87072178)
(1.894944148,1.66079118) (2.439626698,2.91980096)
(-4.828281798,-7.2809754) (-1.359386763,-3.51956976)
(-0.133183214,-1.58376738) (3.722010881,3.14375891)
(-2.062586337,-5.11654085) (-1.593566798,0.43408856)
(-2.247443969,-2.13645461) (2.521617769,1.94467079)
(0.410173496,0.55379156) (0.510253992,0.52654032)
(0.01429945,-0.84184789) (1.531296265,1.9236045)
(1.329605092,2.00210859) (6.134253773,3.58592072)
(1.289002319,2.37691526)

\psset{linewidth=0.3pt,linestyle=solid,linecolor=black}

\psline(-9,-9)(-9,-8.85)    \psline(-9,7)(-9,6.85)
\psline(-9,-9)(-8.85,-9)    \psline(8.5,-9)(8.35,-9)
\psline(-7,-9)(-7,-8.85)    \psline(-7,7)(-7,6.85)
\psline(-9,-7)(-8.85,-7)    \psline(8.5,-7)(8.35,-7)
\psline(-5,-9)(-5,-8.85)    \psline(-5,7)(-5,6.85)
\psline(-9,-5)(-8.85,-5)    \psline(8.5,-5)(8.35,-5)
\psline(-3,-9)(-3,-8.85)    \psline(-3,7)(-3,6.85)
\psline(-9,-3)(-8.85,-3)    \psline(8.5,-3)(8.35,-3)
\psline(-1,-9)(-1,-8.85)    \psline(-1,7)(-1,6.85)
\psline(-9,-1)(-8.85,-1)    \psline(8.5,-1)(8.35,-1)
\psline(1,-9)(1,-8.85)  \psline(1,7)(1,6.85) \psline(-9,1)(-8.85,1)
\psline(8.5,1)(8.35,1) \psline(3,-9)(3,-8.85) \psline(3,7)(3,6.85)
\psline(-9,3)(-8.85,3)  \psline(8.5,3)(8.35,3)
\psline(5,-9)(5,-8.85) \psline(5,7)(5,6.85) \psline(-9,5)(-8.85,5)
\psline(8.5,5)(8.35,5) \psline(7,-9)(7,-8.85) \psline(7,7)(7,6.85)
\psline(-9,7)(-8.85,7) \psline(8.5,7)(8.35,7)

\psline(-9,-9)(-9,7)(8.5,7)(8.5,-9)(-9,-9)

\psline[linestyle=dashed](-9,-9)(7,7)

\rput(-10.1,-9){$-9$}   \rput(-9,-10.1){$-9$}
\rput(-10.1,-5){$-5$}   \rput(-5,-10.1){$-5$}
\rput(-10.1,-1){$-1$}   \rput(-1,-10.1){$-1$}
\rput(-10.1,3){$3$} \rput(3,-10.1){$3$}
\rput(-10.1,7){$7$} \rput(7,-10.1){$7$}

\rput(0.5,-11.5){$y$} \rput(-11.5,0.5){$\hat y_{sp}$}

\rput(-0.5,9.5){(b)}

\end{pspicture}
 \begin{pspicture}(-12.5,-12)(7.5,11)
     \put(0,0){}

\psset{linecolor=black}

\psdots[dotsize=3pt] (0.98636205,2.0012569) (2.509042909,2.8886783)
(7.023502804,5.7410256) (2.496513133,2.4871005)
(2.386586264,2.5932865) (-5.740675877,-6.9854903)
(-1.764218965,-2.2996209) (-0.168430224,-1.0764838)
(4.132001604,3.8349737) (-1.468491645,-2.5395087)
(-0.598098158,0.89597095) (-3.169107714,-2.944895)
(1.918636123,1.740274) (0.999535919,1.0262214)
(1.114740974,1.4107358) (-0.197564846,0.10796947)
(2.401995863,2.4075129) (1.03903931,1.4381808)
(6.94197029,5.1357092) (2.03024014,2.8276385)
(1.054519578,1.3680008) (-4.021225072,-4.018641)
(-1.477331183,-1.9747196) (0.092594171,-0.61911417)
(2.612309542,3.126458) (3.575348652,3.4527071)
(3.953401796,4.9327735) (3.068540578,2.170026)
(0.556268895,0.33401224) (-3.982184083,-4.431828)
(-2.686513892,-1.8471221) (-1.927609008,-1.4490205)
(5.808369411,5.4390637) (0.013804769,-0.10188645)
(0.955361808,1.1235751) (-0.735410796,-0.63379064)
(3.078524452,2.838779) (0.594201807,-0.50423338)
(-0.287440413,-1.0371142) (-1.122338401,-2.0470566)
(8.019167629,6.2225283) (-1.838962635,-2.1587938)
(-1.366758466,-0.68427635) (-3.428010951,-4.3928682)
(2.94024641,2.8851664) (-1.276526235,-1.5924955)
(-0.446418637,-0.81423251) (1.157187722,0.97753338)
(2.065472516,2.7337147) (1.957173202,2.2983475)
(6.947895432,6.325626) (3.406953915,3.478848)
(1.766533864,1.8233858) (-6.401371923,-6.7115481)
(-2.374883527,-2.5654907) (0.289934545,0.61262621)
(0.253073491,1.2636849) (2.530687687,2.8934176)
(5.903635804,4.7541804) (1.178745048,1.129253)
(0.158271769,0.09096172) (-0.610312931,-0.62670283)
(-0.091980244,-0.056593375) (0.465933931,0.33936334)
(3.936377998,3.9628406) (-0.244276121,-0.28700368)
(1.144622722,1.1508464) (0.746007152,0.7825133)
(2.375258447,2.3856695) (0.662890147,0.56569284)
(0.20301397,0.11195611) (0.139660524,0.37424906)
(0.243301625,1.0980539) (2.735884388,2.9080271)
(3.734350536,3.623752) (1.85562568,1.0893129)
(1.196478858,1.0273728) (-3.876484646,-5.6327965)
(-2.903623197,-1.8027144) (-1.29726748,0.40263948)
(3.406995366,3.9475031) (0.292707579,0.44315649)
(1.732847233,2.0236926) (-2.348218294,-2.4838679)
(3.211222771,2.8933691) (-0.241034646,-0.86323895)
(-0.078886725,-0.53573807) (-0.921139955,-1.7680368)
(6.984706961,5.5103542) (-1.256230372,-1.8241698)
(-0.624375784,0.12223157) (-0.259081599,-0.77291551)
(1.090164001,0.90535007) (1.746482933,1.8955338)
(1.404141701,2.5778103) (0.094841936,0.90485626)
(0.421530038,0.960006) (1.422791844,1.7069039)
(4.321491363,3.3714141) (0.157816738,1.7076466)
(2.967198657,3.3175353) (-4.724050459,-4.4917389)
(-1.008550627,-1.7493629) (-0.57786028,-1.6326793)
(7.043218124,5.2806943) (-2.475010141,-3.2671906)
(-1.938584127,-0.7170761) (0.382636754,0.16224187)
(1.74300292,1.5875214) (0.589461673,0.87890215)
(0.746312126,1.4150614) (-0.1661808,0.33757548)
(1.833225075,1.9643645) (0.88363748,1.1523178)
(4.468444574,3.588342) (2.419371571,3.3042843)
(1.805128462,2.0128007) (-5.958874978,-5.7108602)
(-1.830865837,-2.3394948) (0.345894513,0.1279055)
(0.849117234,2.0078912) (3.180286212,3.3840485)
(5.598611727,5.2440234) (1.637564775,0.46330257)
(0.540844407,0.30049065) (-3.066289367,-4.1814729)
(-2.477777397,-1.0022602) (-2.726958328,-1.5504586)
(4.012631686,4.9087887) (0.711781033,0.86112122)
(2.976311458,2.5841523) (-0.546220809,-0.36166486)
(2.87567613,2.6736939) (1.001052531,0.2475432)
(0.013209181,-0.46987981) (0.127480725,0.92549805)
(-0.020469401,0.57135954) (1.837003615,2.0562654)
(3.277153158,2.7624473) (2.604853802,2.4816418)
(0.48168212,0.43131101) (-3.211700058,-3.2126076)
(-1.730019501,-1.6538905) (1.824675028,1.7194578)
(2.483373404,2.8639502) (-0.067224663,0.47717092)
(1.007572388,1.0012107) (-1.980248294,-1.7968161)
(2.555394694,2.2940937) (0.392538173,0.20088383)
(0.464926847,0.44220771) (-1.692126034,-2.0312261)
(7.061558866,5.9712344) (-0.54285416,-0.79603212)
(0.050358572,0.42824397) (-5.035808685,-6.0415946)
(1.890295761,2.0263482) (0.009245474,-0.065205956)
(0.449520913,0.31950092) (0.63806668,0.20946604)
(2.072065701,2.9193888) (2.888544722,3.0455326)
(5.81203582,5.8800187) (2.026481387,2.7098812)
(2.678743101,2.8288814) (-6.403146004,-6.4017711)
(-1.560897268,-2.1640854) (0.037651842,-0.35032439)
(1.657796084,2.2311565) (3.111468315,3.0981853)
(3.058884586,3.8981529) (3.672622572,2.9899725)
(1.123103706,1.0069495) (-5.378737925,-6.0717126)
(-2.959998085,-2.4223178) (-0.481326817,0.53632363)
(4.260300717,4.2366187) (0.027773293,0.078787269)
(0.804052368,1.0890426) (-1.867122514,-2.5265544)
(3.568751699,3.3412091) (-1.766812592,-2.3414216)
(-0.680408822,-1.3673199) (-0.138551395,-0.85559684)
(4.559803651,4.2835343) (-0.717650767,-1.2469083)
(0.695601241,1.2555613) (-4.127539272,-3.9407549)
(1.01380554,0.92737398) (2.425683987,2.5060583)
(2.459114663,3.0425862) (0.27720255,0.63120897)
(1.130918016,1.7577602) (1.85330371,2.2885022)
(7.073458682,5.2691069) (0.050518953,1.6329014)
(1.621914451,2.38031) (-2.675802157,-3.062849)
(-0.861085865,-1.6434798) (-1.658171163,-2.2882855)
(6.856982115,6.3654751) (-0.48529501,-0.52928048)
(0.013559086,0.21575923) (-3.847056381,-3.8754355)
(2.451102067,2.2787006) (0.329231898,0.10169071)
(0.560207864,0.45188718) (0.620206016,0.71495383)
(1.905714462,2.3802919) (1.578982723,2.0253904)
(8.010538802,6.1907095) (3.092290334,2.6886742)
(1.728393941,1.7259324) (-5.386013631,-6.8251682)
(-2.630756624,-2.5249824) (0.126166106,0.97060019)
(-0.032092887,0.55404317) (1.733847948,1.9595475)
(3.292990795,2.7374878) (2.638533837,2.4217211)
(0.45578341,0.36320368) (-3.227564932,-3.2436245)
(-1.827668891,-1.6509987) (-1.133039297,-1.2459505)
(6.126340018,5.2807348) (-0.425795869,-0.7276084)
(0.437111502,0.70929043) (1.877128052,1.7637541)
(2.390423569,2.7581431) (0.952835186,0.95904727)
(0.086330545,-0.021215679) (-1.40169088,-2.0843713)
(5.923524268,6.2223266) (-0.220471792,-0.21409512)
(0.28556768,0.2954213) (-2.402988133,-2.224855)
(2.520477658,2.2651523) (0.410437931,0.2213721)
(0.521061918,0.50642727) (0.858210966,0.73423424)
(1.685668844,2.4918633) (2.148481313,2.5247185)
(7.337205117,6.2241085) (3.178926874,3.1772487)
(2.013292113,2.103087) (-6.276922271,-6.9860628)
(-2.240737704,-2.522256) (0.100169823,1.0516849)
(0.081245395,0.62011845) (1.311254167,1.5289395)
(3.263035819,2.6504644) (2.427632425,1.8716765)
(0.348874686,0.13910762) (-2.962656922,-3.1521339)
(-2.022222115,-1.4485584) (-0.344468213,-0.31472012)
(4.981237283,4.5880613) (-0.286595974,-0.46920411)
(0.814704267,0.92266952) (1.699346543,1.6010318)
(2.237238781,2.4366902) (0.661330336,0.71539706)
(0.161300522,0.13766598) (0.011552507,-0.7011098)
(2.677465641,3.0553713) (3.531741837,3.3837306)
(3.41627771,4.5424583) (3.489893177,2.7351927)
(0.737935757,0.56240961) (-4.673243957,-5.0532198)
(-2.818973957,-2.1660006) (-0.443257368,1.1387163)
(3.324870522,3.6765033) (0.170747998,0.34774383)
(1.07040523,1.5880314) (-2.750196532,-3.2264744)
(3.255716336,3.0144732) (-0.846355669,-1.3903513)
(-0.291437633,-0.77936547) (-0.047204077,-0.8858083)
(3.745735697,3.7754681) (-0.958636608,-1.7442468)
(0.433917625,1.5045512) (-4.342610684,-4.0766924)
(1.490311519,1.3905412) (1.648874893,1.7133596)
(1.756443135,2.0919818) (0.178601765,0.76173501)
(0.601859698,1.2561995) (1.825127149,2.2065374)
(5.805136645,4.3847984) (-0.250102182,1.5160729)
(2.392550754,3.0212964) (-3.472495789,-3.5704474)
(-0.894788693,-1.6857471) (0.115620494,0.66962438)
(0.145439324,0.75359994) (2.190580893,2.3587036)
(3.107831041,2.8669731) (2.159696517,2.3905508)
(0.568607204,0.7345951) (-2.776481395,-2.8622779)
(-1.234732487,-1.5068216) (1.293142007,1.270831)
(3.119538056,3.3901609) (-0.121736993,0.13698987)
(1.091293248,1.0597115) (-0.31106951,-0.19970384)
(2.431215141,2.2570478) (0.417736521,0.35660409)
(0.360590039,0.39798124) (-1.313273527,-1.4234768)
(6.337341491,5.4346677) (-0.42467819,-0.71921907)
(0.37674171,0.67131464) (-4.893078315,-6.6121781)
(1.437493391,1.721765) (-0.160545927,-0.15869142)
(0.338191835,0.3097807) (0.855523576,0.67181766)
(1.698624598,2.5536793) (2.298556983,2.6420521)
(7.076037444,6.1868836) (3.043816571,3.1766004)
(2.118395488,2.2267691) (-6.403405222,-6.9429224)
(-2.113537391,-2.4776164) (-0.979707078,-1.9275462)
(7.58462706,5.8046927) (-1.664943071,-2.1960387)
(-1.175468666,-0.33812702) (1.023500483,0.39422519)
(1.236320145,1.0602159) (1.177351379,1.4488755)
(1.002376528,2.1435572) (-0.324770473,-0.015815169)
(2.484404095,2.4329152) (0.901078736,1.2551125)
(6.003805131,4.5759095) (2.480543465,3.1970015)
(1.199418491,1.4175793) (-4.889664196,-4.7446195)
(-1.785116505,-2.2057429) (0.23008762,0.29469367)
(0.299173535,1.3993314) (3.025601961,3.2496731)
(4.885654324,4.5009998) (1.488318774,0.49235956)
(0.833006175,0.621337) (-3.373254829,-4.8695743)
(-2.746528298,-1.2566835) (-2.452545728,-1.7924162)
(5.28413659,5.4719749) (0.381230568,0.39176223)
(1.722304662,1.6763808) (-0.474868289,-0.2637126)
(2.834785675,2.6692994) (1.419589376,0.34221613)
(-0.124011672,-0.78115085) (0.070885806,0.021716679)
(0.880409878,1.5584533) (2.757022507,2.8367885)
(2.988197274,3.4559929) (3.251855066,2.5538554)
(1.322495148,1.2189605) (-5.112299656,-6.3414774)
(-2.962672124,-2.3816767) (-2.36166389,-0.81506087)
(3.579680172,4.4380142) (0.558749175,0.71485594)
(2.813040182,2.5936173) (-1.265627518,-1.1310039)
(3.037483221,2.730388) (0.459793492,-0.19528012)
(0.047841816,-0.37183725) (-0.530084695,-1.235324)
(5.668954483,4.8564215) (-0.836381328,-1.3809953)
(0.222959247,0.78231067) (-2.600895683,-2.7199545)
(0.956652274,0.81672792) (2.404694354,2.4620506)
(2.150708995,3.0878808) (-0.031597642,0.69748078)
(1.081169014,1.3972596) (0.9435566,1.1618105)
(3.766241868,3.0674561) (1.737439562,2.8312751)
(2.409210064,2.6350398) (-6.058231438,-5.7816763)
(-1.530672792,-2.1669385) (-1.671663019,-2.2583178)
(7.376713221,6.2876794) (-0.736243411,-0.90850198)
(-0.265740929,0.12144266) (-4.522731323,-5.0396556)
(2.332433352,2.3015025) (-0.008596806,-0.20664784)
(0.376706395,0.19684906) (1.740498122,1.6135948)
(2.408920373,2.8241681) (1.494326218,1.8013594)
(6.060759272,6.2482599) (3.565372908,3.5817251)
(1.269998369,1.2779521) (-5.915848772,-5.9412234)
(-2.498674536,-2.5218195) (0.186269534,0.78836959)
(-0.035836854,0.79505188) (2.254536733,2.5597014)
(4.589637352,3.7479391) (1.968472437,1.8064815)
(0.291976512,0.21002785) (-1.979523865,-2.0159415)
(-1.121937585,-1.0086102) (0.370605162,-0.12190236)
(3.55446153,3.7827402) (-0.399640295,-0.62603236)
(1.396683568,1.6056361) (0.348903987,0.48338887)
(2.471160099,2.525421) (1.337002629,0.84117336)
(0.074640673,-0.30298927) (-0.376890608,-1.3892537)
(5.780440593,4.5855088) (-2.218223931,-3.2639999)
(-1.695183759,-0.2233504) (-1.116762296,-1.0610388)
(1.953068261,1.7579309) (0.673893403,0.8229545)
(0.817273109,1.2634209) (0.25074521,0.53842058)
(1.692737308,2.1040146) (1.56220409,2.0083557)
(7.58624883,5.5949791) (0.763552209,1.9828527)
(1.188628957,1.8258525) (-2.736974038,-3.0909657)
(-0.945348152,-1.6448566) (0.358679851,-0.094006949)
(1.557616273,2.5615203) (3.309534123,3.4097376)
(5.322276902,5.4311991) (2.073240146,0.92709691)
(0.437026259,0.18353592) (-3.094954445,-3.9069979)
(-2.416620743,-1.1721237) (-1.146902168,-0.4404266)
(5.102757293,4.8251493) (-0.037229531,-0.12020484)
(0.858928508,1.0525482) (-1.279737017,-1.5986406)
(3.507900492,3.2328581) (-1.117976968,-1.9014601)
(-0.546598359,-1.3164665) (0.080914643,1.0916467)
(0.290122544,0.57478862) (0.975225966,1.0839987)
(1.7731432,1.5428099) (3.219893901,3.1147481)
(0.755684854,0.71285062) (-4.572419177,-4.5349094)
(-2.268532317,-2.1747804) (0.637171698,1.2395561)
(3.772192722,3.7870601) (-0.054472268,0.098402643)
(0.669646956,0.8555524) (1.572692755,1.3445059)
(1.969489004,1.9229019) (0.229635577,0.54535447)
(0.410957258,0.84457007) (-1.809896269,-2.2862754)
(7.129071243,6.2543624) (-0.497381311,-0.634595)
(0.02925847,0.34238804) (-4.751851583,-5.1524559)
(2.179467314,2.155939) (0.273655425,0.11699825)
(0.605286841,0.4673683) (0.336409526,0.56895273)
(0.408382319,1.4461325) (2.563965042,2.9365745)
(6.263818793,5.0425071) (1.894944148,1.791004)
(2.439626698,2.8028392) (-4.828281798,-6.6393211)
(-1.359386763,-2.0441601) (-0.133183214,-0.88743665)
(3.722010881,3.4006958) (-2.062586337,-3.5013317)
(-1.593566798,0.20802865) (-2.247443969,-2.0648139)
(2.521617769,2.2645833) (0.410173496,0.22797265)
(0.510253992,0.50196055) (0.01429945,0.51170699)
(1.531296265,1.8249648) (1.329605092,1.7070116)
(6.134253773,4.5601583) (1.289002319,2.6371888)

\psset{linewidth=0.3pt,linestyle=solid,linecolor=black}

\psline(-9,-9)(-9,-8.85)    \psline(-9,7)(-9,6.85)
\psline(-9,-9)(-8.85,-9)    \psline(8.5,-9)(8.35,-9)
\psline(-7,-9)(-7,-8.85)    \psline(-7,7)(-7,6.85)
\psline(-9,-7)(-8.85,-7)    \psline(8.5,-7)(8.35,-7)
\psline(-5,-9)(-5,-8.85)    \psline(-5,7)(-5,6.85)
\psline(-9,-5)(-8.85,-5)    \psline(8.5,-5)(8.35,-5)
\psline(-3,-9)(-3,-8.85)    \psline(-3,7)(-3,6.85)
\psline(-9,-3)(-8.85,-3)    \psline(8.5,-3)(8.35,-3)
\psline(-1,-9)(-1,-8.85)    \psline(-1,7)(-1,6.85)
\psline(-9,-1)(-8.85,-1)    \psline(8.5,-1)(8.35,-1)
\psline(1,-9)(1,-8.85)  \psline(1,7)(1,6.85) \psline(-9,1)(-8.85,1)
\psline(8.5,1)(8.35,1) \psline(3,-9)(3,-8.85) \psline(3,7)(3,6.85)
\psline(-9,3)(-8.85,3)  \psline(8.5,3)(8.35,3)
\psline(5,-9)(5,-8.85) \psline(5,7)(5,6.85) \psline(-9,5)(-8.85,5)
\psline(8.5,5)(8.35,5) \psline(7,-9)(7,-8.85) \psline(7,7)(7,6.85)
\psline(-9,7)(-8.85,7) \psline(8.5,7)(8.35,7)

\psline(-9,-9)(-9,7)(8.5,7)(8.5,-9)(-9,-9)

\psline[linestyle=dashed](-9,-9)(7,7)

\rput(-10.1,-9){$-9$}   \rput(-9,-10.1){$-9$}
\rput(-10.1,-5){$-5$}   \rput(-5,-10.1){$-5$}
\rput(-10.1,-1){$-1$}   \rput(-1,-10.1){$-1$}
\rput(-10.1,3){$3$} \rput(3,-10.1){$3$}
\rput(-10.1,7){$7$} \rput(7,-10.1){$7$}

\rput(0.5,-11.5){$y$} \rput(-11.5,0.5){$\hat y_{kr}$}

\rput(-0.5,9.5){(c)}

\end{pspicture}

\caption{Smooth supersaturated $\hat y$, spline $\hat y_{sp}$ and
kriging predictions $\hat y_{kr}$ against true simulated values $y$
for the extra design points in Section \ref{ex2}. 
} \label{fig_2d2}

\end{center}
\end{figure}
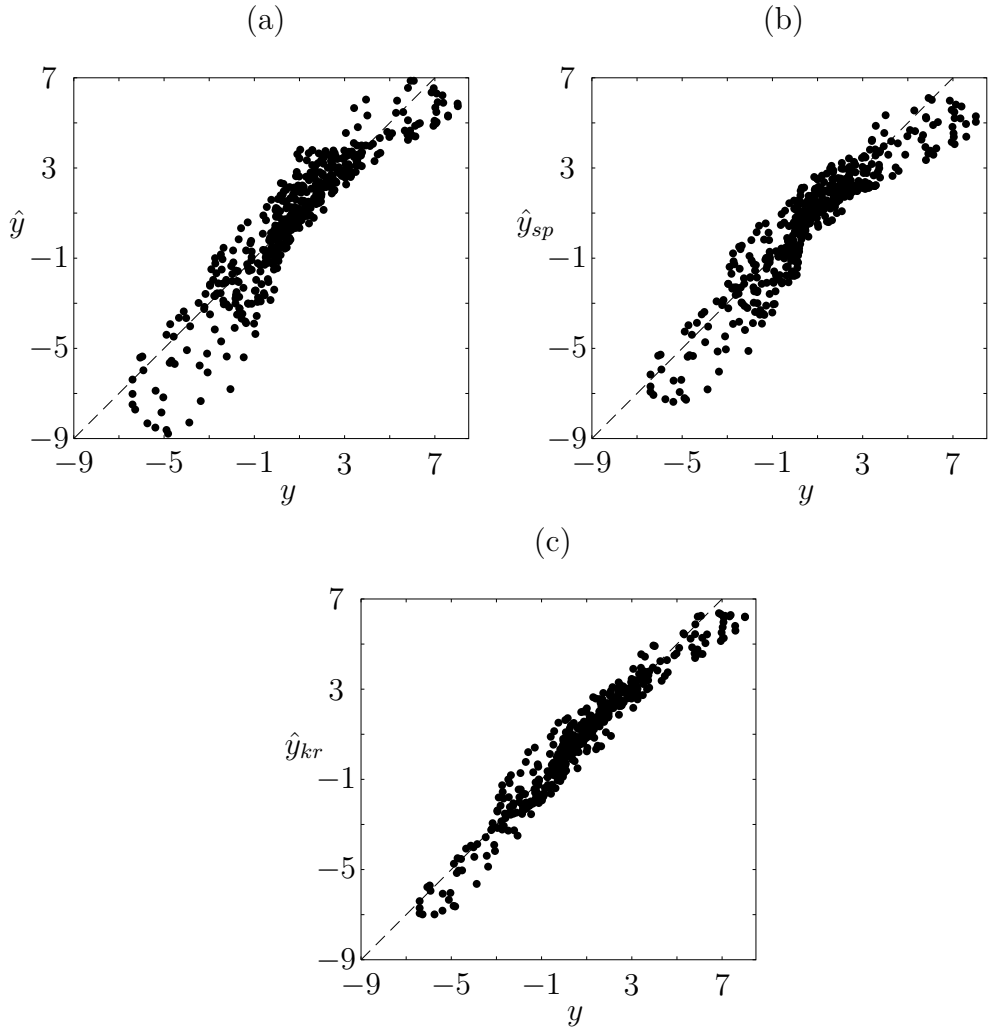

For comparison, a set of predictions were generated for each model
at new design points. The new design points were the next $500$
points from the Sobol' sequence used for the first step. The
predictions obtained with the smooth supersaturated model $\hat y$
are closely correlated with those of the spline $\hat y_{sp}$ and
the kriging $\hat y_{kr}$ models, see Figure \ref{fig_2d} (a) and
(b), only showing bias for low predicted values, especially when
comparing with the kriging model. Additionally, the root mean square
error (RMSE) was computed using the true values $g(x_1,x_2)$ and the
predictions for each of the three models at the extra design points.
The values of RMSE for the smooth supersaturated, spline and kriging
models are $1.117,1.009,0.640$, respectively. This figures represent
the $7.7\%,7.0\%$ and $4.4\%$ of the response range, respectively.
The results show that the smooth supersaturated models are a good
alternative to splines for interpolation, which can also be seen in
Figure \ref{fig_2d2} against the simulated response.

\section{From interpolators to statistical models}
\subsection{Designs points versus knots}
The bulk of the development in this paper concerns the use of the
smooth function as interpolators. However they can be used as
statistical models in a straightforward way. Recall that the
solution are of the form
$$\hat{y}(x) = \hat{\theta}^Tf(x) = y^T B f(x)$$
for the matrix $B$, in one of the equivalent forms in the
development. We see that $\hat{y}(x)$ is linear in the observations
$y$. The idea is to make $y$ a free parameter, that is to change the
role of $y$. Indeed we could relabel $y$ as $\phi$ and write the
model as
$$\hat{y} = \phi^T Bf(x)$$
The design point in $D_n$ become {\em knots} and we are
parameterizing the model by the values at the knots. This is
somewhat familiar in splines. With this change we are free to fit
the models using any regression, stepwise regression, penalised
method etc we choose. There is no requirement to observe at the
knots. But when we have carried out the fitting and write $\hat{y}$
instead of we have the level of smoothness achieved by replacing $y$
by $\hat{y}$ in our formula for $\Psi_2$. Moreover we are free to
choose the location of the knots and the ``real" experimental design
at which to observe. In terms of the dummy design method, this
amounts to a double-dummying: once for the knots and once for the
smoothness; even before we actually take observations.

The function $k(x)=Bf(x)$ can be considered as special kernels each
with a value unity at a design point and zero at other design points
and we can write the model as $\sum_i k_i(x) y_i$ when the $y_i$ are
observations or, in the parametric case just described, as $\sum_i
k_i(x) \phi_i$.

\subsection{Optimal design: for estimation or smoothness}

We restrict the discussion to the case that $K$ is non-singular,
again for simplicity. Then
$$\Psi_2^* = y^T Q y = y^T
 (XK^{-1}X^T)^{-1} y$$
We first note that the design $D_n$, via the design model matrix
$X$, affects the value of the smoothness in the interpolation case,
even without any statistical considerations. Given that  we have to
choose the design {\em before} we observe $y$ one may consider that
some measure of the size of $Q= (XK^{-1}X^T)^{-1}$ may be important.
We may borrow criteria from the optimal design of experiments and
seek to minimize some function of $Q$. In the case that $K$ is
non-singular $\mbox{det}(Q)$ may be used, but as pointed out, since
$K$ is not typically full rank, nor is $Q$.

We consider a small example. Let $n=3,N=5$ and $d=1$ and take the
saturated basis as $1,x,x^2,x^3,x^4$ and let both the design
interval and the integration interval be $\mathcal X $ be $[-1,1]$ .
We need to minimize $\Psi_2 = y^T Q y$ with respect to the choice of
 four design points in $[-1,1]$. After some analysis it can be shown
 that the optimal design take the form $\{-1,-a,a,1\}$ for some
 positive $a$. As expected, because of the two linear terms, the
 matrix $Q$ has rank two. The largest eigenvalue of $Q$ takes the value
 $$\frac{12(1+a^2)}{a^2(1-2a^2+a^4)} $$
 Minimisation of the largest eigenvalue of $Q$ leads to an optimal value of
 $a=1/2\sqrt{-3+\sqrt{17}}\approx0.52988$. Minimising the product of the
 eigenvalues of $Q$ gives $a\approx0.40570$.

In the case that the design $D_{(n)}$ becomes a set of knots we are
free to choose the actual design points separately. If we fit using
smooth supersaturated models this gives an optimal design problem
with the kernels $\{k_j\}$ given above. Continuing with the above
example and guessing that the $D$-optimal on $[0,1]$ for the
optimally smooth kernels obtained by the first solution takes the
form $\{-1, -b, b, 1\}$ we find that $D$-optimal solution as
$$b= \frac{1}{35}
\sqrt{1925+175\sqrt{17}-35\sqrt{2785+480\sqrt{17}}}\approx0.43402,$$
which can,indeed, be confirm to be the $D$-optimum design by
checking against the Kiefer-Wolfowitz General Equivalence Theorem.
One see that these are not the same as the optimal knots.

But now an attractive possibility arises. Optimal design
experimental design for splines has received some attention in the
literature, but it has been considered a somewhat intractable
problem. Now, given that splines can be found as the limit of
polynomial models it may be considered that optimal design for
splines can be found approximately by taking smooth supersaturated
models with large bases, and using one of a number of optimum design
algorithms to find the (approximate) solution. One exchanges a
problem of handling real splines analytically with that of high
dimensional linear algebra. This will be the subject of further
research.

In the case that we are free to choose the knots and the design
points separately, a conceptually simple approach, then, to carry
out two separate separate optimal ``design" problems one for knot
placement for smoothness, as above,  and a second for, say,
$D$-optimality of the design points.

It becomes conceptually harder if we wish to take into account
smoothness and statistical precision in a joint analysis. One might
seek to minimize some portmanteau criterion with respect to a
simultaneous optimizations over design points and knots. If,
moreover, $\Psi_0$ is a statistical criterion such as from
$D$-optimality, we might take as a criterion some weighted
combination:
$$(1-\lambda)\Psi_0 + \lambda \Psi_2$$
As the $y-$values at the knots are now unknown parameters $\phi_i$,
in a linear model we have that the true smoothness is $\Psi_2 =
\phi^TQ \phi$ is  non-linear in $\phi$.

\section{A case study: Engine Emissions Data}\label{sec_example}
\label{ex3}

The performance of a smooth supersaturated model is evaluated
against a kriging model using the engine emissions data set analysed
in \cite{BGW2003}. This data set comes from a computer experiment
and comprises $48$ observations in five factors $N, C, A, B$ and
$M$. An extra set of $49$ observations is available for validation
purposes. The smooth supersaturated model $\hat y$ is constructed
with $100$ terms fitted to the set of $48$ observations. For this
model, $48$ terms correspond to the good saturated basis proposed in
\cite[Section 6.3]{BGW2003}, and this forms $h(x)$. A set of $22$
terms are added to complement missing terms of total degree three
and then a set of extra $30$ terms of total degree four were added.
All the extra $52$ terms described form $g(x)$ and were added using
a degree lexicographic order. Call $\hat y_{sp}$ and $\hat y_{kr}$
to the spline and kriging models constructed with the first data
set. The kriging model $\hat y_{kr}$ was built with a five
dimensional extension of the covariance structure used in Equation
(\ref{ec_krig}).
\begin{figure}[h!]  
\begin{center}
\psset{unit=0.3mm,linewidth=0.85pt}
 \begin{pspicture}(-30,-35)(180,175)
     \put(0,0){}

\psset{linecolor=black}

\psdots[dotsize=3pt](88.4131,83.32748142)
(82.6659,86.20207441)
(90.2547,92.19441811)
(16.2139,9.589634301)
(77.6039,73.37743668)
(5.93,-0.5180246878)
(77.3543,90.8682325)
(52.584,52.48093822)
(42.2756,39.14474112)
(48.5063,47.22527315)
(105.5873,109.3041353)
(16.0462,21.78585182)
(77.6154,74.26829651)
(51.8942,57.01352827)
(30.9128,30.6781001)
(93.8433,93.13402708)
(8.2713,11.77488164)
(48.6699,50.56751601)
(38.5747,43.37388262)
(29.444,30.83198028)
(14.0927,16.38450379)
(121.217,120.7538211)
(60.1144,66.6874279)
(102.2123,96.10929331)
(95.4646,93.4158215)
(53.434,59.04586073)
(59.9135,52.6825832)
(46.6771,31.43023657)
(53.1072,46.80831957)
(12.3041,10.89229896)
(72.3841,75.78432627)
(18.0421,18.38850927)
(68.0012,67.97587236)
(124.5692,123.37365)
(71.3946,65.44358556)
(84.0515,84.91663225)
(39.0258,40.99857838)
(36.2378,33.68162577)
(26.1845,22.28985959)
(18.302,22.74294618)
(24.7408,31.62628742)
(26.5468,26.4622897)
(38.5186,40.16167568)
(27.4956,33.18074472)
(36.689,35.53653198)
(101.4955,90.50131312)
(24.3495,30.98797755)
(93.9711,95.20898148)
(141.6979,135.5985159)

\psset{linewidth=0.5pt,linestyle=solid,linecolor=black}

\psline(0,-1)(0,0.5)    \psline(0,150)(0,148.5) \psline(-1,0)(0.5,0)    \psline(150,0)(148.5,0)
\psline(30,-1)(30,0.5)  \psline(30,150)(30,148.5)   \psline(-1,30)(0.5,30)  \psline(150,30)(148.5,30)
\psline(60,-1)(60,0.5)  \psline(60,150)(60,148.5)   \psline(-1,60)(0.5,60)  \psline(150,60)(148.5,60)
\psline(90,-1)(90,0.5)  \psline(90,150)(90,148.5)   \psline(-1,90)(0.5,90)  \psline(150,90)(148.5,90)
\psline(120,-1)(120,0.5)    \psline(120,150)(120,148.5) \psline(-1,120)(0.5,120)    \psline(150,120)(148.5,120)
\psline(150,-1)(150,0.5)    \psline(150,150)(150,148.5) \psline(-1,150)(0.5,150)    \psline(150,150)(148.5,150)

\psline(-1,-1)(-1,150)(150,150)(150,-1)(-1,-1)

\psline[linestyle=dashed](-1,-1)(150,150)

\rput(-15,0){$0$}   \rput(0,-15){$0$}
\rput(-15,30){$30$} \rput(30,-15){$30$}
\rput(-15,60){$60$} \rput(60,-15){$60$}
\rput(-15,90){$90$} \rput(90,-15){$90$}
\rput(-15,120){$120$}   \rput(120,-15){$120$}
\rput(-15,150){$150$}   \rput(150,-15){$150$}

\rput(75,-30){$\hat y_{sp}$} \rput(-30,75){$\hat y$}
\rput(75,165){(a)}

\end{pspicture}
 \begin{pspicture}(-40,-30)(160,175)
     \put(0,0){}

\psset{linecolor=black}

\psdots[dotsize=3pt]
(92.033596,83.32748142)
(87.90282,86.20207441)
(93.882391,92.19441811)
(11.124812,9.589634301)
(77.268836,73.37743668)
(2.863833,-0.5180246878)
(84.307936,90.8682325)
(48.90309,52.48093822)
(37.450052,39.14474112)
(43.330836,47.22527315)
(108.20592,109.3041353)
(19.757785,21.78585182)
(77.365284,74.26829651)
(49.224028,57.01352827)
(28.786948,30.6781001)
(94.790238,93.13402708)
(11.581712,11.77488164)
(40.203316,50.56751601)
(42.947791,43.37388262)
(23.904384,30.83198028)
(12.186777,16.38450379)
(118.40838,120.7538211)
(65.077333,66.6874279)
(101.09549,96.10929331)
(95.257118,93.4158215)
(52.429644,59.04586073)
(56.202585,52.6825832)
(36.779309,31.43023657)
(50.295476,46.80831957)
(8.4764581,10.89229896)
(73.459724,75.78432627)
(17.527889,18.38850927)
(68.37448,67.97587236)
(123.01676,123.37365)
(68.244275,65.44358556)
(90.74398,84.91663225)
(41.811352,40.99857838)
(31.271562,33.68162577)
(25.894761,22.28985959)
(20.927554,22.74294618)
(28.496826,31.62628742)
(25.085793,26.4622897)
(35.406373,40.16167568)
(32.027104,33.18074472)
(38.962168,35.53653198)
(97.895234,90.50131312)
(26.940354,30.98797755)
(96.71469,95.20898148)
(141.83404,135.5985159)

\psset{linewidth=0.5pt,linestyle=solid,linecolor=black}

\psline(0,-1)(0,0.5)    \psline(0,150)(0,148.5) \psline(-1,0)(0.5,0)    \psline(150,0)(148.5,0)
\psline(30,-1)(30,0.5)  \psline(30,150)(30,148.5)   \psline(-1,30)(0.5,30)  \psline(150,30)(148.5,30)
\psline(60,-1)(60,0.5)  \psline(60,150)(60,148.5)   \psline(-1,60)(0.5,60)  \psline(150,60)(148.5,60)
\psline(90,-1)(90,0.5)  \psline(90,150)(90,148.5)   \psline(-1,90)(0.5,90)  \psline(150,90)(148.5,90)
\psline(120,-1)(120,0.5)    \psline(120,150)(120,148.5) \psline(-1,120)(0.5,120)    \psline(150,120)(148.5,120)
\psline(150,-1)(150,0.5)    \psline(150,150)(150,148.5) \psline(-1,150)(0.5,150)    \psline(150,150)(148.5,150)

\psline(-1,-1)(-1,150)(150,150)(150,-1)(-1,-1)

\psline[linestyle=dashed](-1,-1)(150,150)

\rput(-15,0){$0$}   \rput(0,-15){$0$}
\rput(-15,30){$30$} \rput(30,-15){$30$}
\rput(-15,60){$60$} \rput(60,-15){$60$}
\rput(-15,90){$90$} \rput(90,-15){$90$}
\rput(-15,120){$120$}   \rput(120,-15){$120$}
\rput(-15,150){$150$}   \rput(150,-15){$150$}

\rput(75,-30){$\hat y_{kr}$} \rput(-30,75){$\hat y$}

\rput(75,165){(b)}

\end{pspicture}

\caption{Smooth supersaturated predictions ($\hat y$) against spline
($\hat y_{sp}$) and kriging predictions ($\hat y_{kr}$) for the
validation data set of Section \ref{ex3}.} \label{fig_5d}

\end{center}
\end{figure}
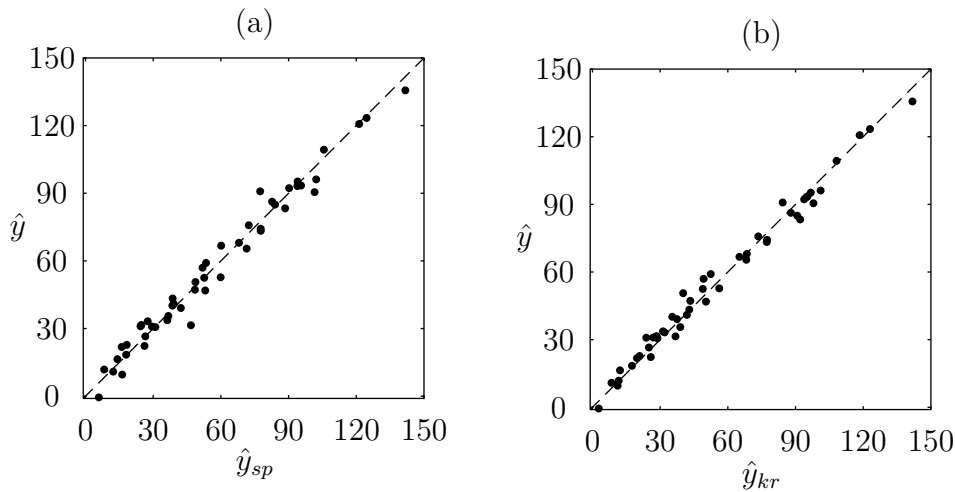

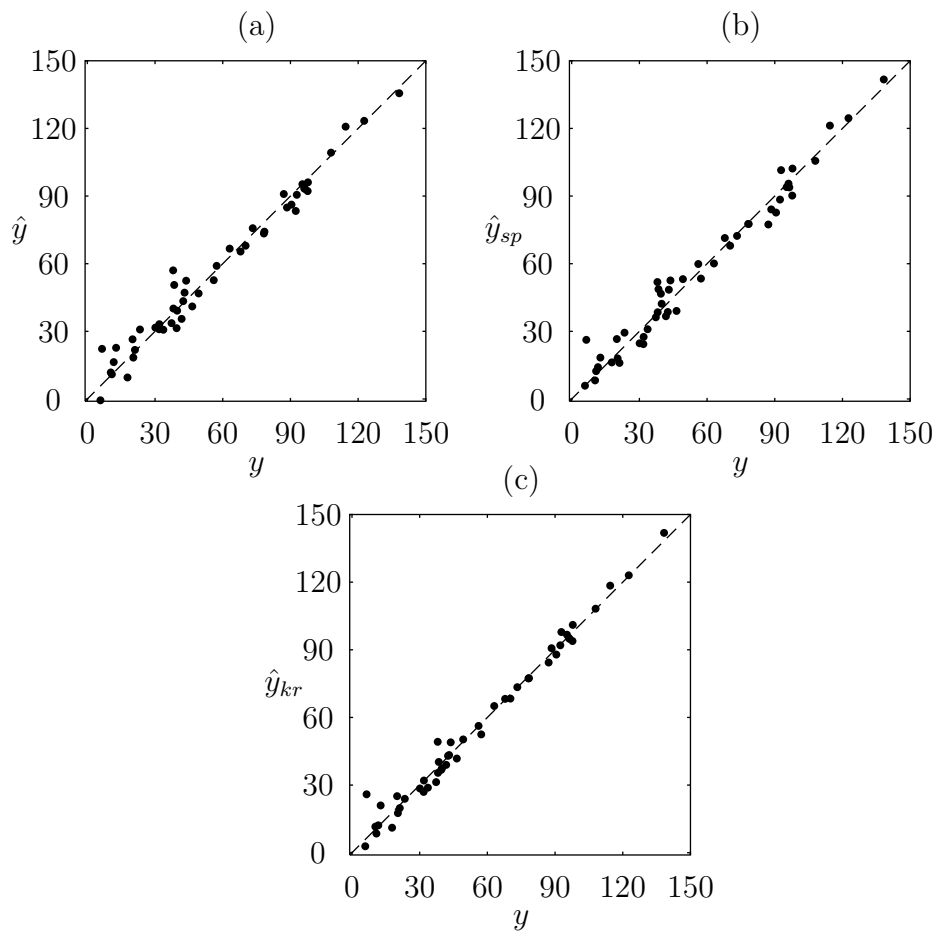
\begin{figure}[h!]  
\begin{center}
\psset{unit=0.3mm,linewidth=0.85pt}
 \begin{pspicture}(-10,-25)(180,175)
     \put(0,0){}

\psset{linecolor=black}

\psdots[dotsize=3pt] (92.288,83.32748142) (90.514,86.20207441)
(97.689,92.19441811) (17.768,9.589634301) (78.216,73.37743668)
(5.851,-0.5180246878) (87.092,90.8682325) (43.738,52.48093822)
(39.822,39.14474112) (43.078,47.22527315) (107.93,109.3041353)
(21.13,21.78585182) (78.482,74.26829651) (37.979,57.01352827)
(33.619,30.6781001) (96.426,93.13402708) (10.36,11.77488164)
(38.507,50.56751601) (42.544,43.37388262) (23.373,30.83198028)
(11.633,16.38450379) (114.45,120.7538211) (63.053,66.6874279)
(97.797,96.10929331) (96.078,93.4158215) (57.251,59.04586073)
(56.049,52.6825832) (39.5,31.43023657) (49.342,46.80831957)
(10.853,10.89229896) (73.301,75.78432627) (20.326,18.38850927)
(70.176,67.97587236) (122.7,123.37365) (67.875,65.44358556)
(88.459,84.91663225) (46.504,40.99857838) (37.266,33.68162577)
(6.5045,22.28985959) (12.699,22.74294618) (30.173,31.62628742)
(20.041,26.4622897) (38.141,40.16167568) (31.904,33.18074472)
(41.796,35.53653198) (92.814,90.50131312) (31.744,30.98797755)
(95.232,95.20898148) (138.25,135.5985159)

\psset{linewidth=0.5pt,linestyle=solid,linecolor=black}

\psline(0,-1)(0,0.5)    \psline(0,150)(0,148.5) \psline(-1,0)(0.5,0)    \psline(150,0)(148.5,0)
\psline(30,-1)(30,0.5)  \psline(30,150)(30,148.5)   \psline(-1,30)(0.5,30)  \psline(150,30)(148.5,30)
\psline(60,-1)(60,0.5)  \psline(60,150)(60,148.5)   \psline(-1,60)(0.5,60)  \psline(150,60)(148.5,60)
\psline(90,-1)(90,0.5)  \psline(90,150)(90,148.5)   \psline(-1,90)(0.5,90)  \psline(150,90)(148.5,90)
\psline(120,-1)(120,0.5)    \psline(120,150)(120,148.5) \psline(-1,120)(0.5,120)    \psline(150,120)(148.5,120)
\psline(150,-1)(150,0.5)    \psline(150,150)(150,148.5) \psline(-1,150)(0.5,150)    \psline(150,150)(148.5,150)

\psline(-1,-1)(-1,150)(150,150)(150,-1)(-1,-1)

\psline[linestyle=dashed](-1,-1)(150,150)

\rput(-15,0){$0$}   \rput(0,-15){$0$}
\rput(-15,30){$30$} \rput(30,-15){$30$}
\rput(-15,60){$60$} \rput(60,-15){$60$}
\rput(-15,90){$90$} \rput(90,-15){$90$}
\rput(-15,120){$120$}   \rput(120,-15){$120$}
\rput(-15,150){$150$}   \rput(150,-15){$150$}

\rput(75,-30){$y$} \rput(-30,75){$\hat y$} \rput(75,165){(a)}

\end{pspicture}
 \begin{pspicture}(-30,-25)(160,175)
     \put(0,0){}

\psset{linecolor=black}

\psdots[dotsize=3pt] (92.288,88.4131) (90.514,82.6659)
(97.689,90.2547) (17.768,16.2139) (78.216,77.6039) (5.851,5.93)
(87.092,77.3543) (43.738,52.584) (39.822,42.2756) (43.078,48.5063)
(107.93,105.5873) (21.13,16.0462) (78.482,77.6154) (37.979,51.8942)
(33.619,30.9128) (96.426,93.8433) (10.36,8.2713) (38.507,48.6699)
(42.544,38.5747) (23.373,29.444) (11.633,14.0927) (114.45,121.217)
(63.053,60.1144) (97.797,102.2123) (96.078,95.4646) (57.251,53.434)
(56.049,59.9135) (39.5,46.6771) (49.342,53.1072) (10.853,12.3041)
(73.301,72.3841) (20.326,18.0421) (70.176,68.0012) (122.7,124.5692)
(67.875,71.3946) (88.459,84.0515) (46.504,39.0258) (37.266,36.2378)
(6.5045,26.1845) (12.699,18.302) (30.173,24.7408) (20.041,26.5468)
(38.141,38.5186) (31.904,27.4956) (41.796,36.689) (92.814,101.4955)
(31.744,24.3495) (95.232,93.9711) (138.25,141.6979)

\psset{linewidth=0.5pt,linestyle=solid,linecolor=black}

\psline(0,-1)(0,0.5)    \psline(0,150)(0,148.5) \psline(-1,0)(0.5,0)    \psline(150,0)(148.5,0)
\psline(30,-1)(30,0.5)  \psline(30,150)(30,148.5)   \psline(-1,30)(0.5,30)  \psline(150,30)(148.5,30)
\psline(60,-1)(60,0.5)  \psline(60,150)(60,148.5)   \psline(-1,60)(0.5,60)  \psline(150,60)(148.5,60)
\psline(90,-1)(90,0.5)  \psline(90,150)(90,148.5)   \psline(-1,90)(0.5,90)  \psline(150,90)(148.5,90)
\psline(120,-1)(120,0.5)    \psline(120,150)(120,148.5) \psline(-1,120)(0.5,120)    \psline(150,120)(148.5,120)
\psline(150,-1)(150,0.5)    \psline(150,150)(150,148.5) \psline(-1,150)(0.5,150)    \psline(150,150)(148.5,150)

\psline(-1,-1)(-1,150)(150,150)(150,-1)(-1,-1)

\psline[linestyle=dashed](-1,-1)(150,150)

\rput(-15,0){$0$}   \rput(0,-15){$0$}
\rput(-15,30){$30$} \rput(30,-15){$30$}
\rput(-15,60){$60$} \rput(60,-15){$60$}
\rput(-15,90){$90$} \rput(90,-15){$90$}
\rput(-15,120){$120$}   \rput(120,-15){$120$}
\rput(-15,150){$150$}   \rput(150,-15){$150$}

\rput(75,-30){$y$} \rput(-30,75){$\hat y_{sp}$}

\rput(75,165){(b)}

\end{pspicture} \begin{pspicture}(-30,-25)(160,175)
     \put(0,0){}

\psset{linecolor=black}

\psdots[dotsize=3pt] (92.288,92.033596) (90.514,87.90282)
(97.689,93.882391) (17.768,11.124812) (78.216,77.268836)
(5.851,2.863833) (87.092,84.307936) (43.738,48.90309)
(39.822,37.450052) (43.078,43.330836) (107.93,108.20592)
(21.13,19.757785) (78.482,77.365284) (37.979,49.224028)
(33.619,28.786948) (96.426,94.790238) (10.36,11.581712)
(38.507,40.203316) (42.544,42.947791) (23.373,23.904384)
(11.633,12.186777) (114.45,118.40838) (63.053,65.077333)
(97.797,101.09549) (96.078,95.257118) (57.251,52.429644)
(56.049,56.202585) (39.5,36.779309) (49.342,50.295476)
(10.853,8.4764581) (73.301,73.459724) (20.326,17.527889)
(70.176,68.37448) (122.7,123.01676) (67.875,68.244275)
(88.459,90.74398) (46.504,41.811352) (37.266,31.271562)
(6.5045,25.894761) (12.699,20.927554) (30.173,28.496826)
(20.041,25.085793) (38.141,35.406373) (31.904,32.027104)
(41.796,38.962168) (92.814,97.895234) (31.744,26.940354)
(95.232,96.71469) (138.25,141.83404)

\psset{linewidth=0.5pt,linestyle=solid,linecolor=black}

\psline(0,-1)(0,0.5)    \psline(0,150)(0,148.5) \psline(-1,0)(0.5,0)
\psline(150,0)(148.5,0) \psline(30,-1)(30,0.5)
\psline(30,150)(30,148.5)   \psline(-1,30)(0.5,30)
\psline(150,30)(148.5,30) \psline(60,-1)(60,0.5)
\psline(60,150)(60,148.5)   \psline(-1,60)(0.5,60)
\psline(150,60)(148.5,60) \psline(90,-1)(90,0.5)
\psline(90,150)(90,148.5)   \psline(-1,90)(0.5,90)
\psline(150,90)(148.5,90) \psline(120,-1)(120,0.5)
\psline(120,150)(120,148.5) \psline(-1,120)(0.5,120)
\psline(150,120)(148.5,120) \psline(150,-1)(150,0.5)
\psline(150,150)(150,148.5) \psline(-1,150)(0.5,150)
\psline(150,150)(148.5,150)

\psline(-1,-1)(-1,150)(150,150)(150,-1)(-1,-1)

\psline[linestyle=dashed](-1,-1)(150,150)

\rput(-15,0){$0$}   \rput(0,-15){$0$} \rput(-15,30){$30$}
\rput(30,-15){$30$} \rput(-15,60){$60$} \rput(60,-15){$60$}
\rput(-15,90){$90$} \rput(90,-15){$90$} \rput(-15,120){$120$}
\rput(120,-15){$120$} \rput(-15,150){$150$}   \rput(150,-15){$150$}

\rput(75,-30){$y$} \rput(-30,75){$\hat y_{kr}$}

\rput(75,165){(c)}

\end{pspicture}

\caption{True values ($y$) against smooth supersaturated predictions
($\hat y$), spline ($\hat y_{sp}$) and kriging predictions ($\hat
y_{kr}$) for the validation data set of Section \ref{ex3}.}
\label{fig_5d2}

\end{center}
\end{figure}

In the validation stage, predictions at the extra $49$ design points
were built using the three models $\hat y,\hat y_{sp}$ and $\hat
y_{kr}$. The values of RMSE for $\hat y, \hat y_{sp}$ and $\hat
y_{kr}$ are $5.844,5.896$ and $4.450$ respectively, which
respectively represent the $4.4\%,4.5\%$ and $3.4\%$ of the range of
the response values. The smooth supersaturated model $\hat y$
compares well with both spline and kriging. Figure \ref{fig_5d}
shows that the predictions with the smooth supersaturated model are
also closely correlated to those obtained with spline and kriging
models. Figure \ref{fig_5d2} also shows the smooth supersaturated
model to be a good predictor of the true response.

\section{Discussion and further research}\label{sec_disc}

We have tried to show in this paper that the simple idea of
extending a basis in regression and using the free parameters which
that gives to increase smoothness give interpolators which have the
same order of magnitude error as the two main alternative: splines
and kriging. For smaller dimensions not too many additional
additional basis terms are need to give a large decrease in
accuracy. Although there is still work to be done on the theory it
seems clear that one can get arbitrarily close to the theoretically
smoothest functions, namely splines. Moreover this can be achieved
for complex regions of integration  and sets of observation points
(designs), limited only by a rank condition.

There a number of ways in which one can generalise or adapt these
methods, which we discuss briefly.
\begin{enumerate}
\item The same analysis will go through for weighted criteria:
$$
\Psi_2 = \int_{\mathcal X}||H(y(x))||^2 w(x)dx,
$$
where $w(x)$ is a non-negative weight function. This simply changes
the definition of $K$ and $\tilde{K}$.
\item The smoothness criteria we
adopted is one of a number in a wider quadratic class such as
$$
\Psi_1 = \int_{\mathcal X}||\bigtriangleup(y(x))||^2dx,
$$
where $\bigtriangleup(y(x))$ is the gradient vector. Another is the
deviation from a target
$$
\Psi_{0,t} = \int_{\mathcal X}|y(x) - t(x)|^2dx,
$$
and one could have weighted versions of them or even weighted
combinations of different criteria.
\item We have ignored analysis based on building in additional,
more statistical criteria, such as cross-validation to have a trade
off between smoothness and statistical variation. A simple way of
taking this forward would be to consider smooth supersaturated as
adding to the catalogue of kernels which are now studied in many
fields such computer experiments, non-parametric regression,
imagining, machine learning and signal processing. They would be
candidates for analysis using stepwise methods, AIC, BIC, LASSO and
so on.

\item A possible advantage of the kernels we have developed is that
their polynomial nature makes them more tractable than, say, splines
in some circumstances; for example for differentiation in
sensitivity analysis, error propagation or integration.

\item We summarize that given detailed attention
to computational issues, it is possible to develop optimal
experimental designs for the high degree, but smooth, kernel models
which arise from the present methods. As mentioned, this may be a
way of tackling optimal design for complex regions.
\item The same methods can be applied for other bases, for example
Fourier bases in one and higher dimensions. Again as the basis order
gets larger one will tend to the optimal spline-like kernels. For
Fourier bases one can gain smoothness by using higher frequencies,
in seeming, but not actual, contradiction to the Nyquist sample
theorem.

\end{enumerate}

\section{Appendix}\label{sec_app}
\subsection{Appendix 1: solution for $\hat{\theta}_0$ and $\hat{\theta}_1$}
It is possible to use block matrix inverse methods, but they are a
little cumbersome. We first find $\hat{\theta}_0$. Writing the
equations out we have
$$
\begin{array}{rll}
X_0\theta_0+X_1\theta_1 & = & y \\
K\theta_1 - X_1^T \lambda & = & 0 \\
X_0 \lambda & = & 0
\end{array}
$$
Solving for $\lambda$ from the second two equations we have
$$\lambda = (X_1K^{-1}X_1^T + X_0X_0^T)^{-1} X_1\theta_1$$
Using this to eliminate $\theta_1$ from the first equation we have
$$X_0^T(X_1K^{-1}X_1^T + X_0X_0^T)^{-1} X_0 \theta_0 =
X_0^T(X_1K^{-1}X_1^T + X_0X_0^T)^{-1} y,$$ giving
$$\hat{\theta}_0 =(X_0^T(X_1K^{-1}X_1^T + X_0X_0^T)^{-1} X_0)^{-1}
X_0^T(X_1K^{-1}X_1^T + X_0X_0^T)^{-1} y,$$

Writing $y^* = y- X_0 \hat{\theta}_0$ we obtain reduced matrix
equation:
$$
\left[
\begin{array}{lr}
 X_1  & 0 \\
 \tilde{K} & -X_1^T \\
 0 & X_0^T
\end{array}
\right] \left[
\begin{array}{lll}
\theta_1\\
\lambda
\end{array}
\right] = \left[
\begin{array}{lll}
y^* \\
0 \\
0
\end{array}
\right]
$$
Left multiplying by the transpose of the matrix on the left and
inverting we have
\begin{equation}
\hat{\theta}_1 = (X_1^TX_1 + \tilde{K}(I - X_1^T
(XX^T)^{-1}X_1)\tilde{K})^{-1}X_1 y^*
\end{equation}
Note that in the case that $X_0$ and $X_1$ have orthogonal columns
we reduce to the standard form
$$\hat{\theta}_0 = (X_0^T X_0)^{-1} X_0^T y$$
This can be achieved by rewriting the supersaturated basis so that
the terms with degree higher than linear (degree one) are orthogonal
to the linear terms with respect to the design. Of course, the
definition of $\tilde{K}$ should be changed accordingly.

\subsection{Equivalence of forms in the case $K$ nonsingular}
The three forms for $\hat{\theta} = By$ where $B$ is one of the
following:

(i) $B_1 =(X_1^TX_1 + K(I-P)K)^{-1} X^Ty$

(i) $B_2 = K^{-1}(X_{11},X_{12})^TQy$

(ii) $B_3 = X^{-1}{I\choose -A_{22}^{-1}A_{21}}$

To show that $B_1 = B_2$ multiply both by $X_1^TX_1 + K(I-P)K$ and
note that $PX^T =0$ to obtain respectively $X^T$ and $X^T X
K^{-1}X^T Q$. But from the definition of $Q$ and using block the
partition inverse formula we see that that $XK^{-1}X^T = Q^{-1}$ and
we are done (reversing the steps).

To show that $B_2=B_3$ we multiply both by ${X^{-1}}^T K$. Then
$B_2$ gives
\[{X^{-1}}^TKK^{-1}(X_{11},X_{12})^TQQ^{-1}={X^{-1}}^T(X_{11},X_{12})^T={I\choose 0},\]
and $B_3$ gives
\[\begin{array}{rcl}
{X^{-1}}^TKX^{-1}{I\choose -A_{22}^{-1}A_{21}}Q^{-1}&=&A{I\choose
-A_{22}^{-1}A_{21}}Q^{-1}=\left(\begin{array}{cc}A_{11}&A_{12}\\A_{21}&A_{22}\end{array}\right){I\choose
-A_{22}^{-1}A_{21}}Q^{-1}\\
&=&{A_{11}-A_{12}A_{22}^{-1}A_{21}\choose A_{21}-A_{22}A_{22}^{-1}A_{21}}Q^{-1}={A_{11}-A_{12}A_{22}^{-1}A_{21} \choose 0}Q^{-1}={I\choose 0}.\\
\end{array}\]
Again, reversing the steps we obtain our result.

\section*{Acknowledgments}

The first and third authors acknowledge the EPSRC grant
GR/S63502/01, while the second and third authors acknowledge the
EPSRC grant EP/D048893/1 (MUCM project).

\bibliographystyle{apalike}
\bibliography{dummy}

\end{document}